\documentclass[prd,aps,floatfix,nofootinbib,twocolumn,tightenlines,showpacs]{revtex4}
\usepackage{dcolumn}
\usepackage{latexsym}
\usepackage{graphicx}
\usepackage{bm}

\def\ttl#1{{``#1''}}

\def\D{\Delta}

\def\svev#1{\left\langle #1\right\rangle}       
\def\tr{{\rm tr}\,}
\def\Tr{{\rm Tr}\,}

\def\Re{{\rm Re\,}}

\long \def \blockcomment #1\endcomment{}

\def\Eq#1{Eq.~(\ref{#1})}
\def\tbeta{\tilde\beta}

\newcommand{\bee}{\begin{equation}}
\newcommand{\ee}{\end{equation}}
\newcommand{\beea}{\begin{eqnarray}}
\newcommand{\eea}{\end{eqnarray}}

\begin{document}

\title{Mass anomalous dimension in sextet QCD}
\author{Thomas DeGrand}
\affiliation{Department of Physics,
University of Colorado, Boulder, CO 80309, USA}
\author{Yigal Shamir}
\author{Benjamin Svetitsky}
\affiliation{Raymond and Beverly Sackler School of Physics and Astronomy,
  Tel~Aviv University, 69978 Tel~Aviv, Israel\\ }

\begin{abstract}
We extend our previous study of the SU(3) gauge theory with $N_f=2$
flavors of fermions in the sextet representation of color.
Our tool is the Schr\"odinger functional method.  By changing the
lattice action, we push the bulk transition of the lattice theory to
stronger couplings and thus reveal the beta function and the mass
anomalous dimension $\gamma_m$ over a wider range of coupling, out to
$g^2\simeq11$.  Our results are consistent with an infrared
fixed point, but walking is not ruled out.
Our main result
is that $\gamma_m$ never exceeds~0.45, making the model
unsuitable for walking technicolor.
We use a novel method of extrapolation to the large-volume/continuum limit,
tailored to near-conformal theories.
\end{abstract}

\maketitle

\section{Introduction}
For some time we have been studying SU($N$) gauge theories with fermions
in the symmetric two-index representation of
color~\cite{Shamir:2008pb,DeGrand:2008kx,DeGrand:2009hu,DeGrand:2010na,
Svetitsky:2010zd,DeGrand:2011qd,DeGrand:2011vp,DeGrand:2012qa}.
These are among the theories that have been
proposed~\cite{Sannino:2004qp,Hong:2004td,Dietrich:2005jn} as candidate
models for walking
technicolor~\cite{Holdom:1981rm,Yamawaki:1985zg,Appelquist:1997fp,Hill:2002ap}.
Here we present our most recent work on the SU(3) gauge theory with
$N_f=2$ flavors of fermions in the sextet representation.

A technicolor theory must supply Goldstone bosons in order to generate
masses for the weak vector bosons.  To that end, the theory must break
chiral symmetry spontaneously.  In order to generate quark and lepton
masses as well while avoiding large effects of flavor-changing neutral
currents, one demands the additional property of {\em walking\/}.  Here
a large separation between the technicolor scale $\Lambda_{TC}$ and the
``extended'' technicolor scale $\Lambda_{ETC}$ is attained by having a
near-zero of the beta function; the running coupling stalls for many
decades in energy before chiral symmetry breaking sets in at large
distances.  Furthermore, the effect of the large ratio
$\Lambda_{ETC}/\Lambda_{TC}$ on the technifermion condensate has to
be enhanced by a large anomalous dimension $\gamma_m$ of the mass operator
$\bar\psi\psi$.  Current estimates~\cite{Chivukula:2010tn} require
$\gamma_m\simeq1$.
In fact $\gamma_m=1$ emerges from analyses based on the gap equation for walking technicolor~\cite{Holdom:1981rm,Yamawaki:1985zg} (see also \cite{Kaplan:2009kr}).

Since the beta function and $\gamma_m$ are the important ingredients of
walking technicolor, our work has focused on measuring them.
We do this
using Schr\"odinger-functional (SF) techniques
\cite{Luscher:1992an,Luscher:1993gh,Sint:1993un,Sint:1995ch,Jansen:1998mx,
  DellaMorte:2004bc},
which are a lattice implementation of the background-field method.
For other instances of the Schr\"odinger functional applied to technicolor candidates, see~\cite{Appelquist:2007hu,Appelquist:2009ty,Hietanen:2009az,Bursa:2009we,Bursa:2010xn,Hayakawa:2010yn,Karavirta:2011zg}.

Our first effort \cite{Shamir:2008pb} used Wilson's fermion action with
an added clover term~\cite{Sheikholeslami:1985ij} to reduce $O(a)$
effects, and was limited to lattices of linear size $L=4a$, $6a$, and~$8a$.
The result was a discrete beta function that appeared to cross
zero at a renormalized coupling $g^2\simeq2.0$, indicating an infrared
fixed point (IRFP).  An IRFP indicates conformal physics at large
distances, the antithesis of confinement.  Intending to understand (and
reduce) discretization effects, we then~\cite{DeGrand:2010na} went to
larger lattices, $L/a=6$, 8, 12,~16, and began using hypercubic
smearing---fat links~\cite{Hasenfratz:2001hp, Hasenfratz:2007rf}---in
the fermion action.  This work showed that the IRFP of
Ref.~\cite{Shamir:2008pb} was but a lattice artifact.  Thanks to its
numerical stability, the fat-link action also enabled us to simulate at
stronger couplings, out to $g^2\simeq4.6$, corresponding to a bare
coupling $\beta=4.4$.  At stronger bare couplings we encountered a phase
transition that makes it impossible to tune the hopping parameter
$\kappa$ so as to make the quark mass zero.  The result of
Ref.~\cite{DeGrand:2010na}, then, is a beta function that is smaller in
magnitude than the two-loop result but that does not cross zero in the
accessible range of couplings.

In Ref.~\cite{DeGrand:2010na} we also calculated the anomalous dimension
$\gamma_m$ according to the method
of~\cite{Sint:1998iq,Capitani:1998mq,DellaMorte:2005kg,Bursa:2009we}.
We found that $\gamma_m$ first follows the one-loop curve but its rise
slows at strong couplings so that $\gamma_m\alt0.6$ in the range of
couplings that we could reach.

In our work on the SU(4) gauge theory with two-index (decuplet)
fermions~\cite{DeGrand:2011vp,DeGrand:2012qa}, we
encountered the same phase transition that prevents simulation of the
massless theory.  We found that augmenting the pure-gauge part of the
action with a new fat-plaquette term can move this transition to
stronger bare coupling and stronger renormalized coupling as well.  In
the present paper, we present the result of applying this strategy to
the SU(3) theory.  The new action enables us to reach $g^2\simeq11$,
which is in the vicinity of the zero of the two-loop beta function, discussed by
Caswell~\cite{Caswell:1974gg} and by Banks and Zaks~\cite{Banks:1981nn}.

While at first glance our calculated beta function crosses zero near the two-loop zero, 
further analysis shows that this result is not stable under extrapolation to the continuum limit.
It is quite possible that the beta function of this theory approaches zero in the range of coupling that we can study, and then veers away from zero at yet stronger couplings---much as envisioned in scenarios of walking.

Whether the beta function crosses zero or no, we find, in agreement with our earlier work, that the anomalous dimension has left the one-loop curve and leveled off.
The limit we set at our strongest coupling is $\gamma_m\alt 0.45$.  
Our earlier work, based on the pure plaquette gauge action, quoted values approaching 0.6, but this was without attempting to remove lattice artifacts.
We report here an analysis of lattice artifacts that allows smooth extrapolation to the continuum limit.
The results for both actions move downwards and the error bars grow under extrapolation.
As a result, the results for the two actions are not in violent disagreement.
The small value of $\gamma_m$ near the (real or approximate) fixed point
spells trouble for any use of the present theory as walking technicolor.

The plan of this paper is as follows.  In Sec.~\ref{sec:action} we
present the improved lattice action, which is the only difference in the
simulation method between this paper and Ref.~\cite{DeGrand:2010na}.
For other details of our simulations we refer the reader to
Ref.~\cite{DeGrand:2010na}.  We also briefly discuss the ensembles
we generated, and how we deal with the autocorrelations
of our observables.  We proceed in Sec.~\ref{sec:beta} to
present our results for the running coupling and the beta function.
We re-analyze the data of Ref.~\cite{DeGrand:2010na} according to
the lights of our later paper on the SU(2) theory~\cite{DeGrand:2011qd},
where we learned to take advantage of the slow running in order to make
maximum use of the several lattice sizes in play.  Naturally, we add in
the results of the new simulations obtained with the augmented gauge action.
Our first results show an apparent zero in the beta function; this will not survive our analysis of discretization error later in the paper.
Section~\ref{sec:gamma} contains our numerical
results for the mass anomalous dimension $\gamma_m$.
We study the finite-lattice corrections to both $\gamma_m$ and the beta function in Sec.~\ref{sec:syst}.
The former reaches a smooth continuum limit, albeit with larger error bars than those of the raw results  of Sec.~\ref{sec:gamma};
the latter is not so well-behaved.
In Sec.~\ref{sec:summary} we summarize our results and place them in the context of other work.
In the appendix we examine the fermion contribution to the one-loop SF coupling and explain that it does not provide any guidance for the range of couplings and volumes that we study.

\section{Lattice action and ensembles\label{sec:action}}

Our action contains a fermion term and two pure gauge terms.  The
fermion term $\bar\psi D_F \psi$
is the conventional Wilson action,
supplemented by a clover term~\cite{Sheikholeslami:1985ij} with coefficient
$c_{\text{SW}}=1$~\cite{Shamir:2010cq}.
The gauge links in the fermion action
are fat link variables $V_\mu(x)$.  The fat links are the normalized
hypercubic (nHYP) links of Ref.~\cite{Hasenfratz:2001hp}, where for each
link $(x,\mu)$ one takes the weighted average $V_\mu(x)$ of links in
neighboring hypercubes with weights $\alpha_1=0.75$, $\alpha_2=0.6$,
$\alpha_3=0.3$, reunitarized and subsequently promoted to the sextet
representation.

The gauge action is
\begin{eqnarray}
S_G &=&-
\frac{\beta}{2N} \sum_{\mu\ne\nu} \Re \Tr U_\mu(x) U_\nu(x+\hat\mu)
U_\mu^\dagger(x+\hat\nu) U_\nu^\dagger(x)
\nonumber\\
& &- \frac{\beta_6}{2d_f} \sum_{\mu\ne\nu} \Re \Tr V_\mu(x) V_\nu(x+\hat\mu)
V_\mu^\dagger(x+\hat\nu) V_\nu^\dagger(x),
\nonumber\\
\label{eq:gaugeaction}
\end{eqnarray}
wherein the first term is the usual sum of plaquettes of fundamental
thin link variables, while the second term contains plaquettes made of
fat links in the sextet representation as in the fermion action.
$N=3$ is the number of colors while $d_f=6$ is the dimension of the fermion
representation, here the sextet.
The weak-coupling expansion of $S_G$ gives the effective bare coupling~\cite{DeGrand:2011vp},
\begin{equation}
  \frac{1}{g_0^2} = \frac{\beta}{2N} + \frac{T_6 \beta_6}{6} \,,
\label{g0eff}
\end{equation}
where $T_6=5/2$ is the group trace in the fermions' representation.%
\footnote{
  We presented a test of weak-coupling universality in our paper on the SU(4) theory~\cite{DeGrand:2012qa}.
}

As before, we employ the hybrid Monte Carlo (HMC) algorithm in our
simulations.  The molecular dynamics integration is accelerated with an
additional heavy pseudo-fermion field as suggested by
Hasenbusch~\cite{Hasenbusch:2001ne}, multiple time
scales~\cite{Urbach:2005ji}, and a second-order Omelyan
integrator~\cite{Takaishi:2005tz}.  We determined the critical hopping
parameter $\kappa_c = \kappa_c(\beta)$ by setting to zero the quark mass,
as defined by the unimproved axial Ward identity on lattices of size $L=12a$.

Without a systematic search, we found that choosing
$\beta_6=+0.5$ removes the strong-coupling phase transition
so that we can run at bare couplings down to $\beta=2.0$;
at smaller $\beta$ the acceptance deteriorates rapidly, especially
for larger volumes, so that we did not go far enough to find out
if and where the strong-coupling transition turns up.
At $\beta=2.0$, the running coupling for $L=6$ turns
out to be $g^2\simeq11$.  This is close to the two-loop
Banks--Zaks zero at $g^2=\frac{13}{194}(16\pi^2)\simeq10.6$.
We list in Table~\ref{table:runs} the values of $(\beta,\kappa_c)$ and the
number of trajectories run at each volume, along with the
length of the trajectories and the acceptance.
Poor acceptance forced us to shorten the trajectory length in many cases from the usual value of 1.

The observables we measure are the (inverse)
SF running coupling, $1/g^2$,
and the pseudoscalar renormalization factor, $Z_P$.
(We measure $Z_P$ on the same configurations used to determine $1/g^2$.)
Both of them turn out to have long autocorrelation times.
We monitored and controlled this problem by running 4 or 8 streams in parallel at each $\beta$ and $L$.
After analyzing each stream separately,
we fit the results of
the streams together to a constant.
We demanded that the $\chi^2/{\rm dof}$
of the constant fit not exceed 6/3 for 4 streams, or 10/7 for 8 streams.
For the largest volume $L=16a$ at the strongest coupling $\beta=2.0$,
we were not able to overcome the autocorrelations in $1/g^2$
even with nearly 30,000 trajectories.
We therefore omit this point from the analysis of the running coupling.
The autocorrelations in $Z_P$, on the other hand, did allow a consistent
determination, and thus
we keep this point in the analysis of the mass anomalous dimension.

In this paper we present our new results, obtained with $\beta_6=0.5$, alongside
the $\beta_6=0$ results presented in our earlier paper~\cite{DeGrand:2010na}.
For the more extensive study of discretization error in Sec.~\ref{sec:syst}, we supplemented the data of Ref.~\cite{DeGrand:2010na} with simulations on new lattice sizes $L=10a,14a$ at two values of $\beta$ in strong coupling, as shown in Table~\ref{table:runs0}.

\begin{table}
\caption{$\beta_6=0.5$ ensembles generated at the bare couplings
  $(\beta, \kappa_c)$ for the lattice sizes $L$ used in this study.
  Listed are the total number of trajectories for all streams at given
$(\beta,L)$,
  the trajectory length, and the HMC acceptance.}
\begin{center}
\begin{ruledtabular}
\begin{tabular}{ccrccc}
  $\beta$ &
  $\kappa_c $ &
  $L/a$  &
  trajectories&
  trajectory&
  acceptance\\
&&&(thousands)&length&\\
\hline
3.5 & 0.13349 &  6 & 74.8 & 1.0 & 0.93 \\
            & &  8 & 15.5 & 0.5 & 0.97 \\
            & & 10 & \hspace{.6ex} 8.0 & 1.0 & 0.75 \\
            & & 12 & 37.0 & 0.5 & 0.88 \\[2pt]
2.5 & 0.13991 &  6 & \hspace{.6ex} 8.8 & 1.0 & 0.61 \\
            & &  8 & 14.3 & 1.0 & 0.43 \\
            & & 10 & 20.4 & 0.5 & 0.75 \\
            & & 12 & 35.6 & 0.5 & 0.50 \\
            & & 14 & 11.6 & 0.5 & 0.61 \\
            & & 16 & 17.1 & 0.5 & 0.48 \\[2pt]
2.0 & 0.14273 &  6 & 17.2 & 1.0 & 0.65 \\
            & &  8 & 14.2 & 0.5 & 0.61 \\
            & & 10 & 12.0 & 0.5 & 0.67 \\
            & & 12 & 13.4 & 0.5 & 0.48 \\
            & & 16 & 29.6 & 0.4 & 0.38 \\
\end{tabular}
\end{ruledtabular}
\end{center}
\label{table:runs}
\end{table}
\begin{table}
\caption{$\beta_6=0$ ensembles generated at the new lattice sizes $L=10a$ and~$14a$, which we add to the ensembles listed in Ref.~\cite{DeGrand:2010na}.
  Columns as in Table~\ref{table:runs}.}
\begin{center}
\begin{ruledtabular}
\begin{tabular}{ccrccc}
  $\beta$ &
  $\kappa_c $ &
  $L/a$  &
  trajectories&
  trajectory&
  acceptance\\
&&&(thousands)&length&\\
\hline
4.8 & 0.13173 & 10 & \hspace{.6ex} 9 & 1.0 & 0.95 \\[2pt]
4.4 & 0.13510 & 10 & 16 & 1.0 & 0.80 \\
            & & 14 & 12 & 0.5 & 0.85 \\
\end{tabular}
\end{ruledtabular}
\end{center}
\label{table:runs0}
\end{table}

\section{Beta function \label{sec:beta}}

The computation of the running coupling proceeds exactly as described in
Ref.~\cite{DeGrand:2010na}, with the same boundary conditions on the
fermion and gauge fields.  In brief, one imposes Dirichlet boundary
conditions at the time slices $t=0,L,$ and measures the response of the
quantum effective action.  The coupling emerges from a measurement of
the derivative of the action with respect to a parameter $\eta$ in the
boundary gauge field,
\begin{equation}
 \frac{K}{g^2(L)}  =
  \left.\svev{\frac{\partial S_{G}}{\partial\eta}
  -\tr \left( \frac{1}{D_F^\dagger}\;
        \frac{\partial (D_F^\dagger D_F)}{\partial\eta}\;
            \frac{1}{D_F} \right)}\right|_{\eta=0}  .
            \label{deta}
\end{equation}
The constant $K$ can be calculated
directly from the classical continuum action.
Only $g_0$, \Eq{g0eff}, appears in the latter,
which assures that $K=12\pi$ regardless of $\beta_6$.

\begin{figure}
\begin{center}
\includegraphics[width=.95\columnwidth,clip]{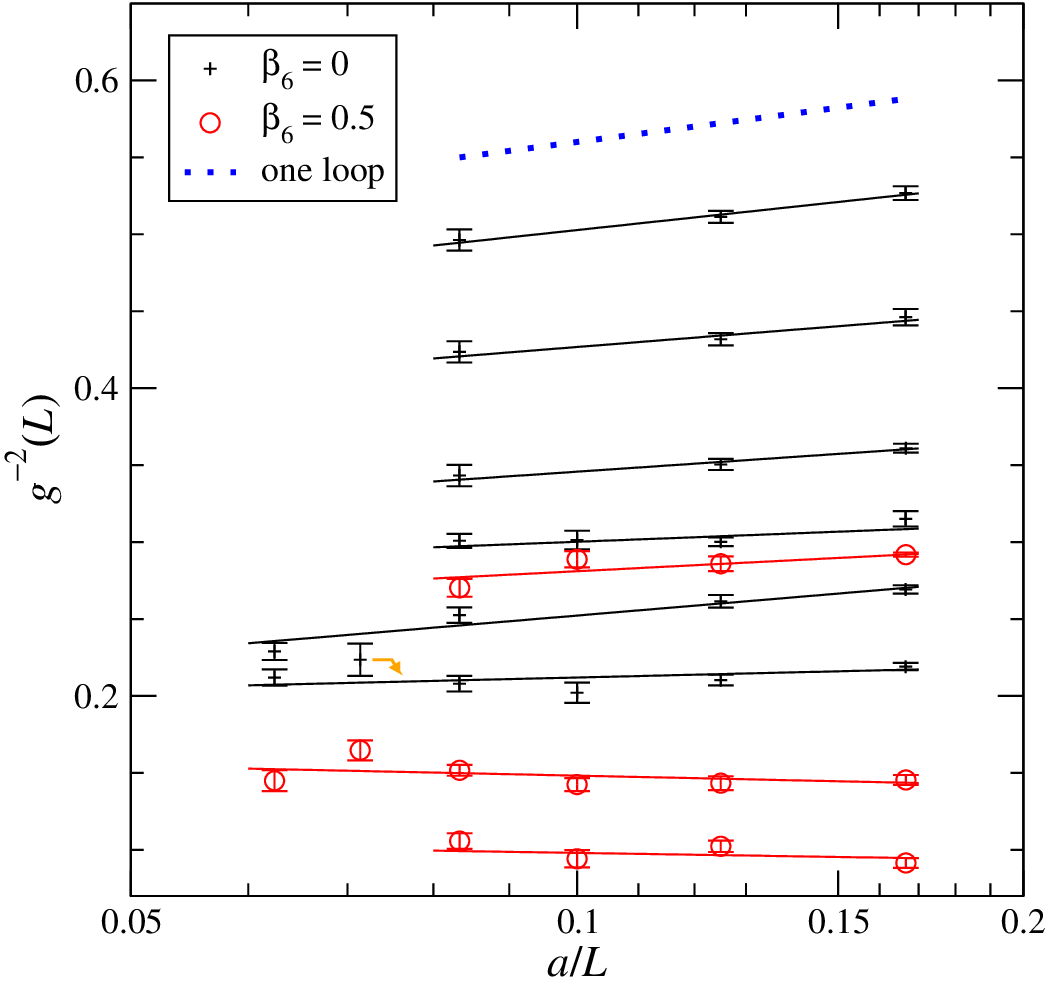}
\end{center}
\caption{
  Running coupling
  $1/g^2$ vs.~$a/L$.  The crosses are from
  simulations with $\beta_6=0$ (Ref.~\cite{DeGrand:2010na} and Table~\ref{table:bf0}): top to bottom, $\beta=5.8,$ 5.4, 5.0, 4.8,
  4.6, and 4.4.  The circles are from simulations with
  $\beta_6=0.5$ (Table~\ref{table:bf05}): top to bottom, $\beta=3.5$,
  2.5, and~2.0.  The straight lines are linear fits~[\Eq{linfit}]
  to each set of points at given $(\beta,\beta_6)$;
  the slope gives the beta function.
  The dotted line shows the expected slope from one-loop running.
\label{fig:couplings}}
\end{figure}

We presented in Ref.~\cite{DeGrand:2010na}
the values of the running coupling $g^2$
for a number of values of $(\beta,\kappa_c)$ with $\beta_6=0$.
Our new results for $\beta_6=0.5$ are shown in Table~\ref{table:bf05}, with supplementary data on new volumes for $\beta_6=0$ shown in Table~\ref{table:bf0}.
Both sets are plotted in Fig.~\ref{fig:couplings}.%
\footnote{We have dropped from consideration the data given in \cite{DeGrand:2010na} for $(\beta,\beta_6)=(4.3,0)$ since these were taken in a metastable state
beyond the first-order boundary.}

\begin{table*}
\caption{Running coupling, \Eq{deta},
evaluated at the bare coupling $(\beta, \kappa_c)$ with $\beta_6=0.5$
on lattices of size $L$.  The omission of the result for $L=16a$ at $\beta=2.0$ is explained
in the text.}
\begin{center}
\begin{ruledtabular}
\begin{tabular}{cllllll}
$\beta$ & \multicolumn{6}{c}{$1/g^2$}\\
\cline{2-7}
    & $L=6a$     & $L=8a$     & $L=10a$    & $L=12a$    & $L=14a$    & $L=16a$    \\
\hline
3.5 & 0.2918(13) & 0.2859(48) & 0.2888(53) & 0.2703(58) & \hfil--    & \hfil--    \\
2.5 & 0.1454(32) & 0.1433(45) & 0.1424(42) & 0.1517(36) & 0.1647(65) & 0.1449(68) \\
2.0 & 0.0915(32) & 0.1023(37) & 0.0942(56) & 0.1057(51) & \hfil--    & \hfil*     \\
\end{tabular}
\end{ruledtabular}
\end{center}
\label{table:bf05}
\end{table*}
\begin{table*}
\caption{Running coupling for the strong-coupling cases with $\beta_6=0$.
Data for $L=10a,14a$ are from the new simulations listed in Table~\ref{table:runs0}, while the other data are from Ref.~\cite{DeGrand:2010na}.
Data for $\beta=4.6$ were taken with $\kappa=\kappa_c=0.13320$.}
\begin{center}
\begin{ruledtabular}
\begin{tabular}{cclllll}
$\beta$ &  \multicolumn{6}{c}{$1/g^2$}\\
\cline{2-7}
    & $L=6a$     & $L=8a$     & $L=10a$    & $L=12a$    & $L=14a$    & $L=16a$  \\
\hline
4.8 & 0.3151(50) & 0.3002(28) & 0.3013(61) & 0.3008(45) & \hfil--    & \hfil--  \\
4.6 & 0.2692(27) & 0.2615(41) & \hfil--    & 0.2525(50) & \hfil--    & 0.2289(56)\\
4.4 & 0.2191(24) & 0.2103(34) & 0.2021(66) & 0.2080(50) & 0.2235(105)& 0.2119(53)\\
\end{tabular}
\end{ruledtabular}
\end{center}
\label{table:bf0}
\end{table*}

It is convenient to define the beta function $\tbeta(u)$ for $u\equiv1/g^2$ as
\bee
  \tbeta(u) \equiv \frac{d(1/g^2)}{d\log L}
  = 2\beta(g^2)/g^4 = 2u^2 \beta(1/u)
\label{invbeta}
\ee
in terms of the conventional beta function $\beta(g^2)$.
\begin{figure*}
\begin{center}
\includegraphics[width=.95\columnwidth,clip]{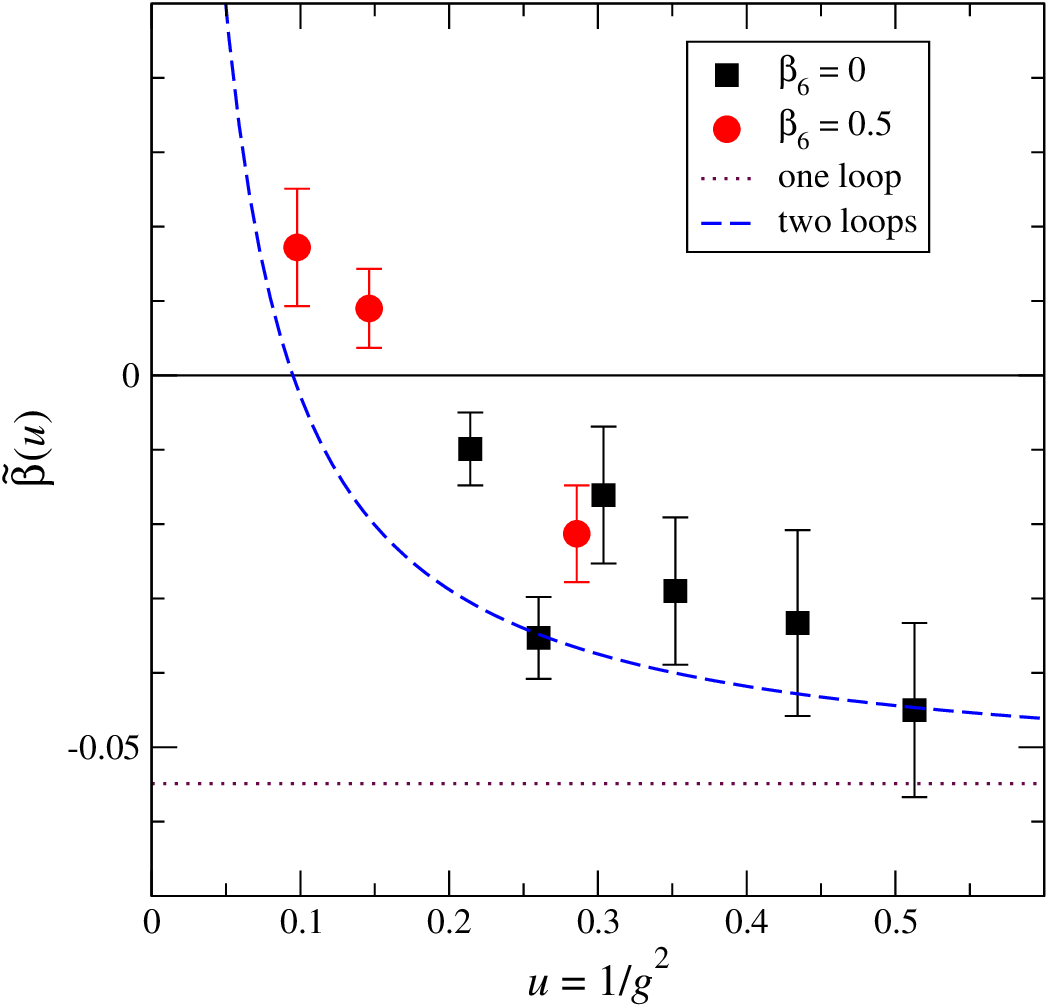}\hfill  
\includegraphics[width=.95\columnwidth,clip]{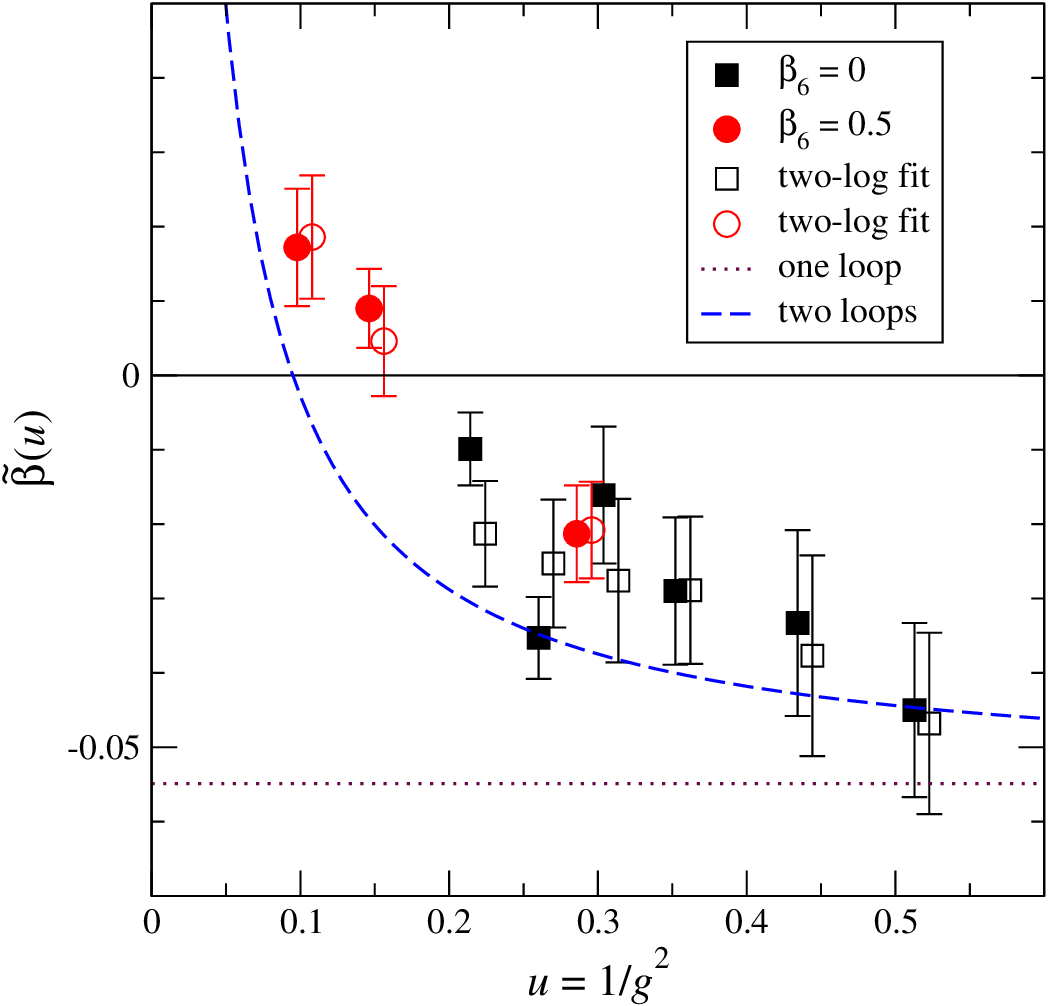}
\end{center}
\caption{Left: Beta function $\tbeta(u)$  plotted as a function of $u(L=8a)$.
The squares are from the $\beta_6=0$ data while the circles
are from $\beta_6=0.5$.
Results are extracted from the
linear fits~(\ref{linfit}), as shown in Fig.~\ref{fig:couplings}.
Plotted curves are the one-loop (dotted line)
and two-loop (dashed line) beta functions.
Right: Filled symbols as on left; empty symbols derive from fits to \Eq{l2fit},
in which a $\log^2$ term has been added.
The empty symbols have been slightly displaced to the right.
No correction has been made for discretization errors.
\label{fig:beta}}
\end{figure*}
As discussed in Ref.~\cite{DeGrand:2011qd}, the slow running of the
coupling suggests extracting the beta function at each $(\beta,\kappa_c)$ from
a linear fit of the inverse coupling
\bee
u(L)=c_0+c_1 \log \frac L{8a}\ .
\label{linfit}
\ee
With this parametrization, $c_0$ gives the inverse coupling $u(L=8a)$,
while $c_1$ is an estimate for the beta function $\tbeta$ at this coupling.

For a first look, we fit the data points for {\em all\/} $L$ to extract the
slopes at the given bare parameters, ignoring discretization errors that must be inherent in the smallest lattices.
These fits are shown in Fig.~\ref{fig:couplings}.  
Values of the beta function $\tbeta(u)$ obtained from these fits
are plotted as a function of $u(L=8a)$ in Fig.~\ref{fig:beta}.
One can see that the results for $\beta_6=0$
and for $\beta_6=0.5$ are consistent with each other.
Also shown are the one- and two-loop approximations from the expansion
\begin{equation}
  \tbeta(u) = -\frac{2b_1}{16\pi^2}
  -\frac{2b_2}{(16\pi^2)^2}\frac1u +\cdots ,
\label{btilde}
\end{equation}
where $b_1=13/3$ and $b_2=-194/3$.

The assumption behind the linear fits is that $\tbeta$
is small so that $u(L/a)$ changes very slowly with the volume;
this behavior is apparent in Fig.~\ref{fig:couplings}.
Nonetheless, where there are four or more volumes to be fitted for a given coupling the $\chi^2$ is not particularly good.
Corrections to the approximate \Eq{linfit} come from discretization
errors, as well as from the slight deviation from constancy of the
continuum beta function over the range of volumes.
We defer consideration of discretization errors to Sec.~\ref{sec:syst}.
Variation of the continuum beta function
gives rise to higher powers of $\log L$.  Adding the next-to-leading
term, at each bare coupling we fit
\bee
u(L)=c_0 +c_1 \log L/8a +c_2 (\log L/8a)^2 .
\label{l2fit}
\ee
From the definition of a beta function it follows that
$c_1$ continues to provide an estimate for $\tbeta$
at $u=1/g^2(L=8a)$.  The results of these fits are shown as empty symbols in Fig.~\ref{fig:beta}.  It is evident that there is only a small change
compared to the linear fits of \Eq{linfit}.
In our analysis of discretization error, given below, we use only the single logarithm.

At the two strongest couplings the beta function plotted in Fig.~\ref{fig:beta} is positive,
indicating the existence of an infrared fixed point.  
As we shall see, this conclusion does not survive consideration of the discretization error.
It is possible that
the beta function remains negative even at the strongest couplings studied.
Hence, we cannot rule out
a ``walking'' scenario wherein the beta function comes close to zero,
but never actually turns positive.

Nevertheless, Figs.~\ref{fig:couplings} and~\ref{fig:beta} demonstrate an important qualitative result, which will be unaltered by any analysis of lattice error: 
The coupling constant runs more slowly than one-loop perturbation theory over the entire range of bare parameters. 
This is in marked contrast to QCD, where the running coupling
runs faster than the one-loop (or the two-loop) beta function as one enters strong coupling~\cite{DellaMorte:2004bc}.

\section{Mass anomalous dimension \label{sec:gamma}}

We derive the mass anomalous dimension from the scaling with $L$ of the
pseudoscalar renormalization factor $Z_P$.
The latter is calculated by taking the ratio
\bee
Z_P = \frac {c \sqrt{f_1}}{f_P(L/2)}.
\label{eq:ZP}
\ee
$f_P$ is the propagator from a wall source at the $t=0$ boundary
to a point pseudoscalar operator at time $L/2$.
The normalization of the wall source is removed by the $f_1$ factor,
which is a boundary-to-boundary correlator.
The constant $c$, which is an arbitrary normalization,
is $1/\sqrt{2}$ in our convention.

\begin{figure}
\begin{center}
\includegraphics[width=.95\columnwidth,clip]{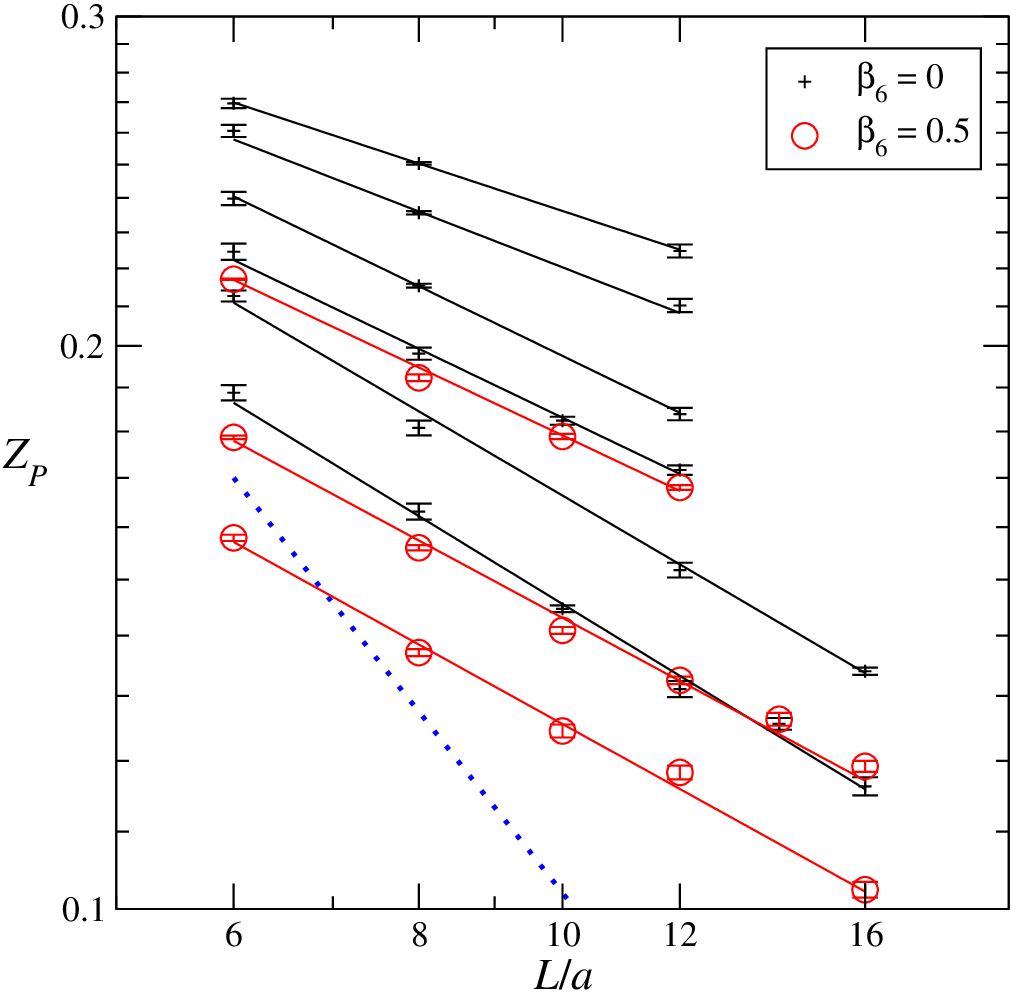}
\end{center}
\caption{The pseudoscalar renormalization constant
  $Z_P$ vs.~$L/a$.  The crosses are from simulations with $\beta_6=0$,
  $\beta=5.8$ to 4.4.  The circles are from simulations with
  $\beta_6=0.5$ (Table~\ref{table:ZP}): top to bottom, $\beta=3.5$, 2.5,
  and~2.0.  The straight lines are linear fits to each set of points at given
  $(\beta,\beta_6)$; the slope gives $-\gamma_m$.
  The hypothetical dotted line corresponds to $\gamma_m=1$.
\label{fig:ZP}}
\end{figure}

We present in Tables~\ref{table:ZP} and~\ref{table:ZP0} the results of calculating $Z_P$
in our runs with $\beta_6=0.5$ and with $\beta_6=0$; we plot them in Fig.~\ref{fig:ZP}.
\begin{table*}
\caption{Pseudoscalar renormalization factor $Z_P$ evaluated at the
  couplings $(\beta,\kappa_c)$, with $\beta_6=0.5$, for lattice
  sizes $L$.}
\begin{center}
\begin{ruledtabular}
\begin{tabular}{cllllll}
$\beta$ 
&\multicolumn{6}{c}{$Z_P$}\\
\cline{2-7}
&   $L=6a$   & $L=8a$       & $L=10a$    & $L=12a$    & $L=14a$    & $L=16a$   \\
\hline
3.5 & 0.2171(2) & 0.1923(8) & 0.1789(6)  & 0.1680(5)  & \hfil--    & \hfil--   \\
2.5 & 0.1787(4) & 0.1560(5) & 0.1409(6)  & 0.1325(6)  & 0.1263(10) & 0.1192(8) \\
2.0 & 0.1579(6) & 0.1371(6) & 0.1245(10) & 0.1183(10) & \hfil--    & 0.1024(10)\\
\end{tabular}
\end{ruledtabular}
\end{center}
\label{table:ZP}
\end{table*}

\begin{table*}
\caption{Pseudoscalar renormalization factor $Z_P$ for the strong-coupling cases with $\beta_6=0$.
Data for $L=10a,14a$ are from the new simulations listed in Table~\ref{table:runs0}, while the other data are from Ref.~\cite{DeGrand:2010na}.}
\begin{center}
\begin{ruledtabular}
\begin{tabular}{cclllll}
$\beta$ &  \multicolumn{6}{c}{$Z_P$}\\
\cline{2-7}
    & $L=6a$     & $L=8a$     & $L=10a$    & $L=12a$    & $L=14a$   & $L=16a$   \\
\hline
4.8 & 0.2246(23) & 0.1981(15) & 0.1824(9)  & 0.1716(10) & \hfil--   & \hfil--   \\
4.6 & 0.2127(14) & 0.1808(16) & \hfil--    & 0.1518(14) & \hfil--   & 0.1340(6) \\
4.4 & 0.1888(18) & 0.1631(16) & 0.1447(6)  & 0.1311(13) & 0.1256(9) & 0.1163(13)\\
\end{tabular}
\end{ruledtabular}
\end{center}
\label{table:ZP0}
\end{table*}

\begin{figure*}
\begin{center}
\includegraphics[width=.95\columnwidth,clip]{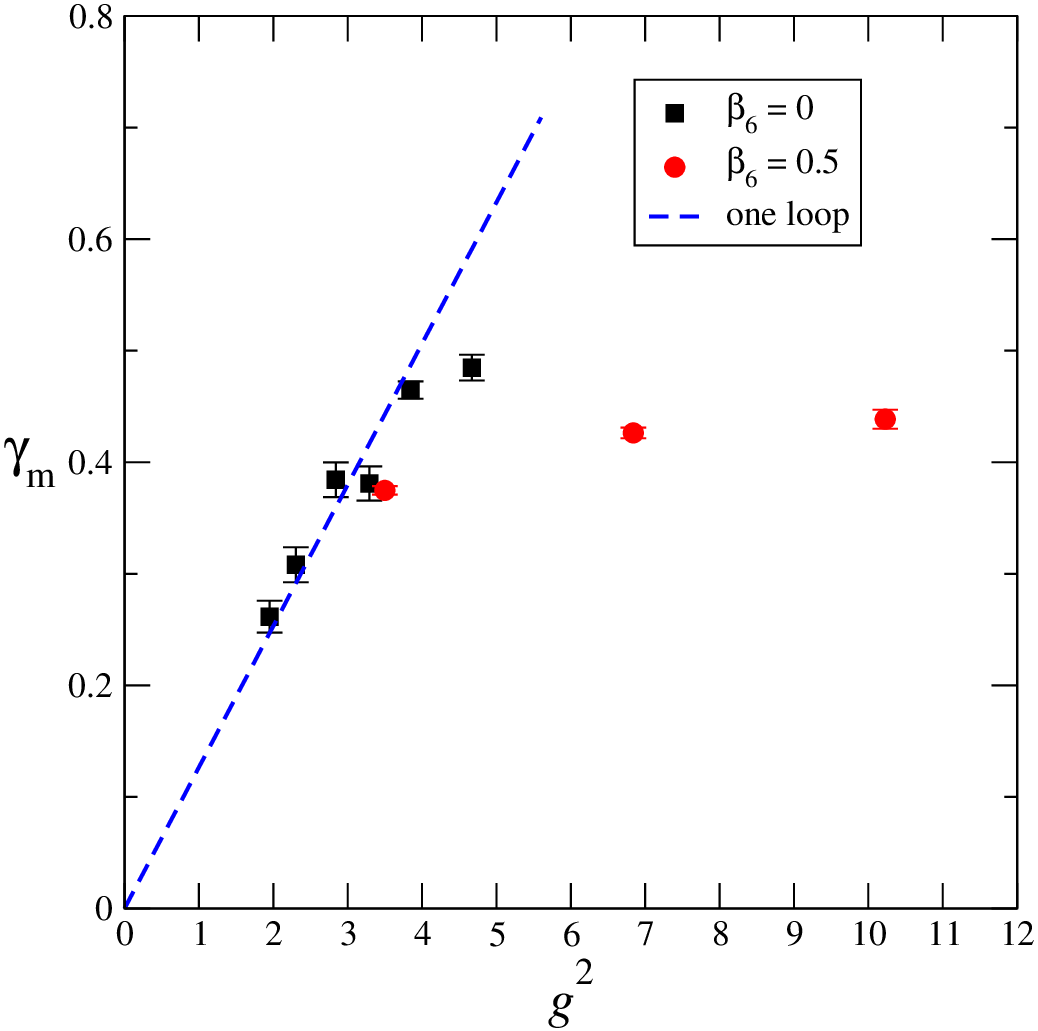}\hfill 
\includegraphics[width=.95\columnwidth,clip]{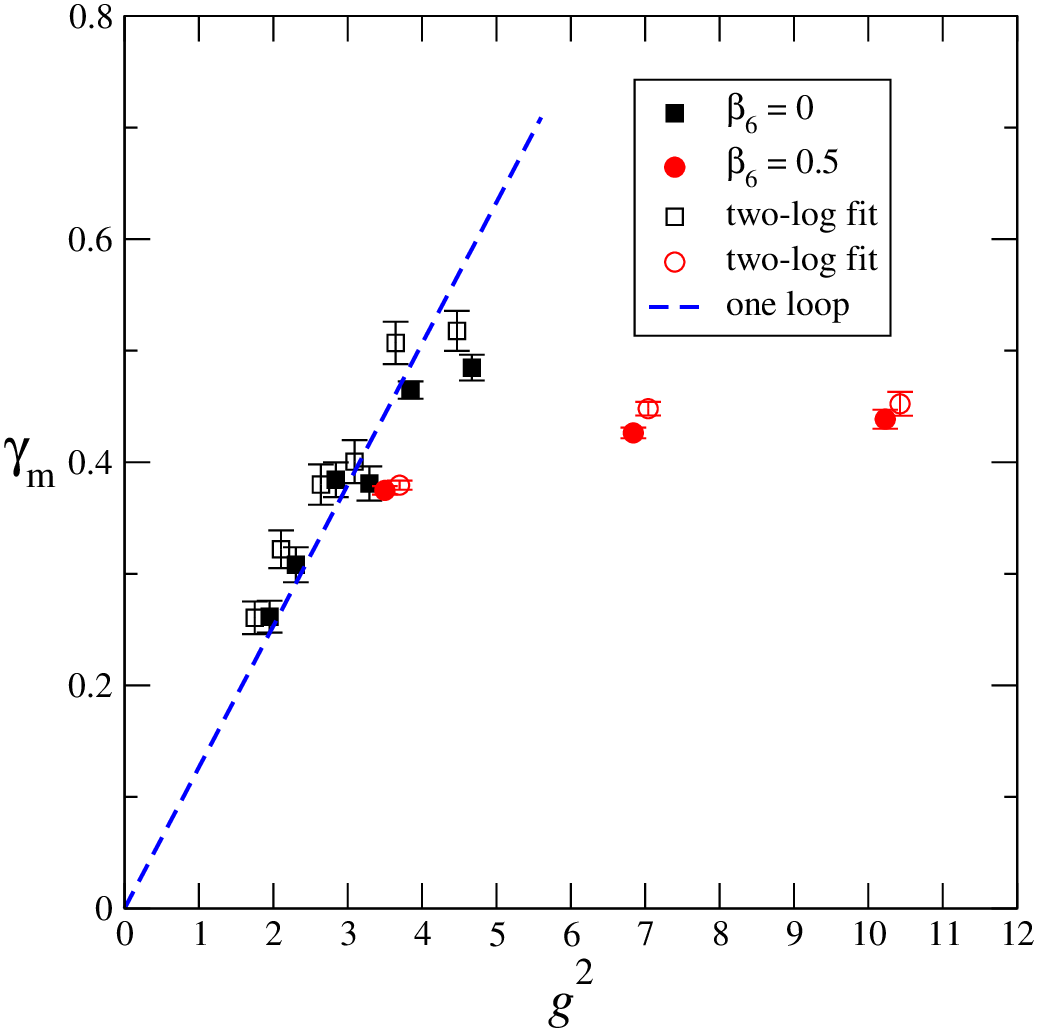}
\end{center}
\caption{Left: Mass anomalous dimension $\gamma(g^2)$ plotted against $g^2(L=8a)$.  
The squares are from the $\beta_6=0$ data while the circles are from $\beta_6=0.5$.
Results are from the linear fits
shown in Fig.~\ref{fig:ZP}.
The line is the one-loop result.
Right: Filled symbols as on left; empty symbols derive from fits to \Eq{l2fitZ},
in which a $\log^2$ term has been added.
The empty symbols have been slightly displaced horizontally.
No correction has been made for discretization errors.
\label{fig:gamma}}
\end{figure*}

The slow running suggests \cite{DeGrand:2011qd} that we may extract $\gamma_m$
by applying the approximate scaling formula
\bee
Z_P(L)=Z_P(L_0)\left(\frac{L_0}L\right)^\gamma ,
\label{Zfp}
\ee
that is, from the slopes of the lines drawn in Fig.~\ref{fig:ZP}.
These linear fits are analogous to \Eq{linfit}:
\bee
\log Z_P(L)=c_0+c_1 \log \frac L{8a} .
\label{linfitZ}
\ee
Again, we begin by fitting to all volumes at a given bare coupling simultaneously, leaving
for later the discretization errors due to the smallest lattices. 
The individual fits are shown as straight lines in Fig.~\ref{fig:ZP}.
The outstanding feature of Fig.~\ref{fig:ZP} is that in no case does the slope of any data
set approach $-1$, meaning that $\gamma_m$ (Fig.~\ref{fig:gamma}) never reaches unity.
This qualitative observation survives all further analysis.

Allowing for running of the coupling, we have also considered
a fit function analogous to \Eq{l2fit},
\bee
\log Z_P(L)=c_0 +c_1 \log L/8a +c_2 (\log L/8a)^2 .
\label{l2fitZ}
\ee
For both fits the mass anomalous dimension
at $g^2(L=8a)$ is given by $-c_1$.
We show the results of both fit types in Fig.~\ref{fig:gamma},
plotted against the running coupling $g^2(L=8a)$.
It is apparent that the result for $\gamma_m(g^2)$ changes little when the fit
is broadened, or in other words, that scaling violations due to the nonzero beta function are small.

A comparison of results for the two lattice actions shows that there is some
disagreement.  
The two strongest-coupling points obtained with
$\beta_6=0$
lie well above the line connecting the $\beta_6=0.5$ points.
As we shall see, analysis of discretization errors reduces this discrepancy.
Moreover, the rough bound $\gamma_m\alt0.5$ will be strengthened by extrapolation to the continuum limit.

\section{Estimating discretization error \label{sec:syst}}
Before showing our extrapolations to the continuum/infinite volume limit, let us make some general comments.

Lattice studies of quantum chromodynamics approach the continuum limit in the weak-coupling regime.
QCD possesses a physical scale $\Lambda$.
A systematic approach to discretization error gives an expansion of physical quantities in powers of $\Lambda a$, with which one may extrapolate to the $a\to0$ limit.
Perturbation theory gives a guide to the coefficients in the expansion.
The $L\to\infty$ limit is taken separately; it is governed by the mass gap $M$ of the theory and the corrections are functions of $ML$.

The theory considered here differs from lattice QCD in several essential features.
In the first place, the coupling runs much more slowly than in QCD.
After all, the theory was selected for study because this is true in the two-loop beta function;  our numerical results bear this out nonperturbatively.
In the second place, we seek the putative infrared fixed point in strong coupling.
Along with a strong SF coupling at the volume scale $L$, then, we have slow running to a still-strong coupling at the lattice scale $a$.
There is a perturbative scale $\Lambda_1$ that emerges from one-loop running, but it never comes into play.
Much as in truly conformal theories, discretization errors are functions of $a/L$---they are indistinguishable from finite-volume corrections. 
It is clear from this that any perturbative estimate of discretization error is pointless.

Nonetheless, we offer a discussion of the fermion contribution to the one-loop SF
coupling in the appendix.  We point out that,
even were we to work at couplings where one-loop perturbation theory is valid,
for our lattices the $O(a^2/L^2)$ corrections are so large that they cannot be disentangled from $O(a/L)$.
This means that any fit function used to extrapolate to $a/L=0$ constitutes a model rather than an expansion about the limit.

Since our main result is the upper bound on the anomalous dimension $\gamma_m$, we begin with the analysis of discretization error in this quantity.
We focus on the strong-coupling data displayed in Tables~\ref{table:ZP} and~\ref{table:ZP0}.
The $\beta_6=0.5$ data show the dramatic departure of $\gamma(g^2)$ from the one-loop line (see Fig.~\ref{fig:gamma}).
Comparison to the $\beta_6=0$ shows an apparent discrepancy between the two actions that begs explanation.
We present several extrapolations to $a/L=0$, which result in a systematic shift 
downwards of the results.
At the same time, the results from the two actions are brought closer to agreement, both through the propagation of statistical error in the extrapolation and through uncertainty surrounding the method of extrapolation to $a/L=0$.

Similar analyses of the beta function are inconclusive.
The dependence of the running coupling $g^2$ on lattice size $L$ is irregular and not amenable to smooth extrapolation to the continuum.
In the end, we have to let our lattice results stand as they are.

\subsection{Anomalous dimension}

The general behavior we expect for $Z_P(L)$ is
\beea
\label{eq:sigPgamma}
  Z_P(L) &=& A \exp\left[-\int^1 \frac {dt}{t}\,
  \gamma_m\left(g^2(tL)\right)\right]   + P_1\frac aL  \nonumber\\
  &&+P_2\left(\frac aL\right)^2
 + \cdots.
\eea
For a slowly-running coupling, the first term simplifies to $A\left(L/a\right)^{-\gamma_m(g^2)}$.
The original procedure \cite{DellaMorte:2005kg} is to extract the step scaling function $\sigma_P(g^2,s)$ from the ratio of $Z_P$ measured on two lattices,
\bee
  \label{eq:sigma_p}
  \sigma_P(v,s) = \left. {\frac{Z_P(sL)}{Z_P(L)}}
  \right|_{g^2(L)=v}.
\ee
For slow running, this is just
\bee
\sigma_P(g^2,s) = s^{ - \gamma_m(g^2)}.
  \label{eq:gamma}
\ee
The traditional analysis%
\footnote{The truly traditional analysis keeps $g^2$ fixed by varying the bare couplings as $L$ is changed.
As we do throughout this paper, we take advantage of the slow running to do our analyses at fixed $(\beta,\beta_6)$.
The variation of $g^2$ is negligible.}
continues by computing $\sigma_P(g^2,s)$ at fixed $s$ for several values of $L$
and extrapolating $L$ to infinity. Let us begin our analysis by doing variations on the traditional fit.

\subsubsection{Extrapolation of the step scaling function \label{ss:Extrap}}

\begin{figure*}[hbt]
\begin{center}
\includegraphics[width=.95\columnwidth,clip]{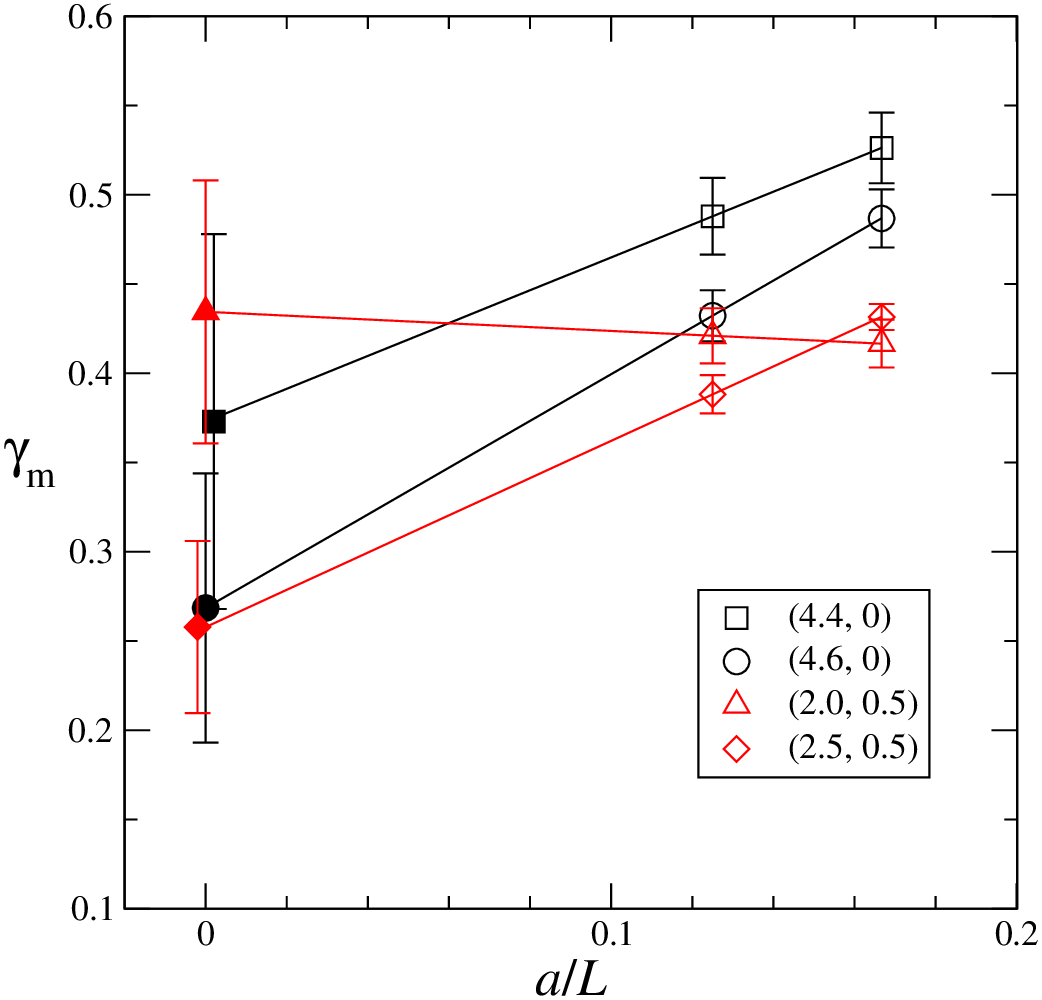}\hfill 
\includegraphics[width=.95\columnwidth,clip]{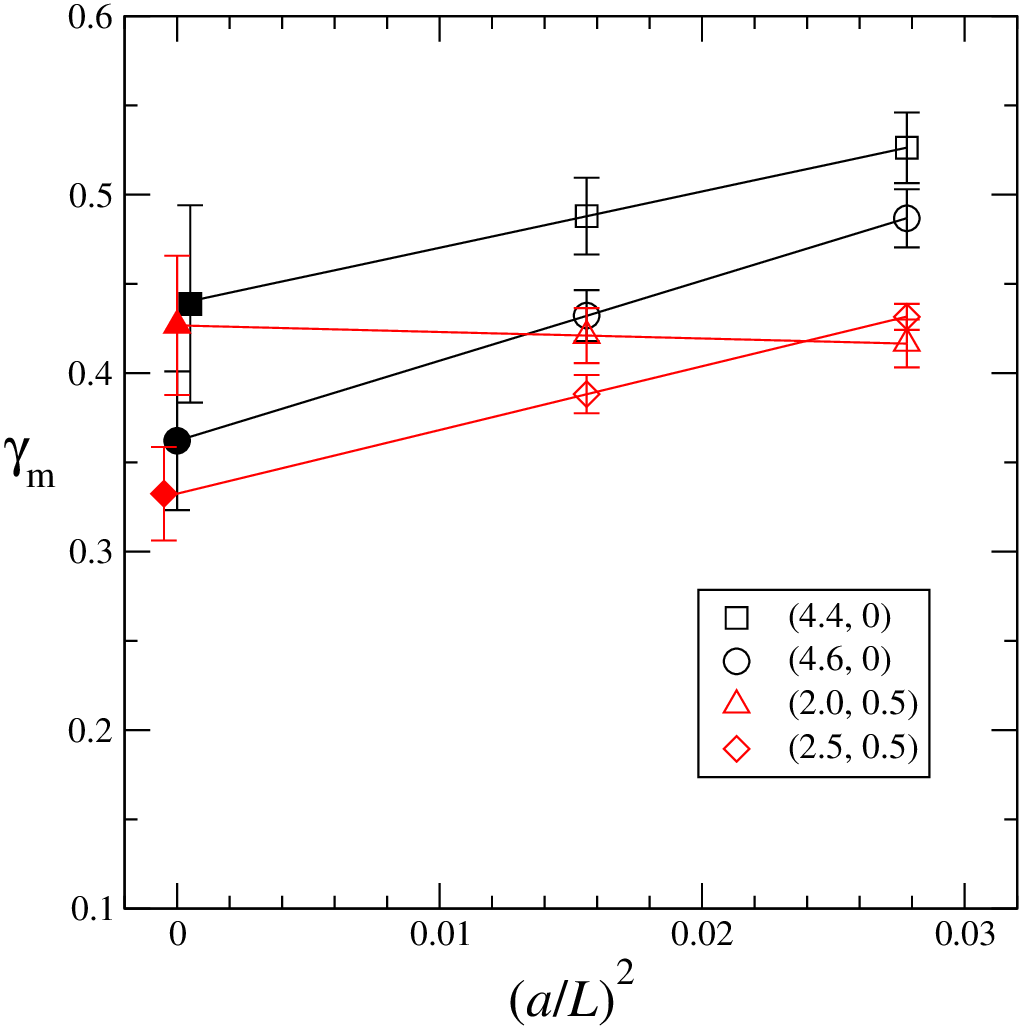}
\end{center}
\caption{$\gamma_m$ from ``traditional'' fits to pairs of lattices $(L,2L)=(6a,12a)$ and $(8a,16a)$, plotted
versus $a/L$ (left) and versus $(a/L)^2$ (right). The extrapolated values are shown near the origin.
Couplings $(\beta,\beta_6)$ are as shown.
\label{fig:tradzp}}
\end{figure*}

At%
\footnote{We sometimes refer to the data sets by their $\beta$ values.  The tables show that there is no confusion if we omit mention of $\beta_6$.}
 $\beta=4.4$, 4.6, and 2.0, 2.5, we can do the most traditional fit of all: fix the ratio $s=2$ and compare
$\sigma_P$ for the pair $L=(6a,12a)$ with $L=(8a,16a)$. The results for 
\bee
\gamma_m(L)=\frac{\log\sigma_P}{\log2}
\ee
are plotted in Fig.~\ref{fig:tradzp} against $L$ of the smaller lattice.

In all four cases,there is a discernible $L$ dependence in the data.
Now we face the problem of how to extrapolate in $L$.
As noted above, we are away from any perturbative limit, and so theory cannot guide us.
Choosing models, we perform extrapolations linearly in $L$,
\bee
\gamma_m(L)= \gamma_m+C\frac aL,
\label{l1a1fit}
\ee
or quadratically,
\bee
\gamma_m(L)= \gamma_m+C\left(\frac aL\right)^2.
\label{l1a2fit}
\ee
The fits, with their extrapolations to $a/L=0$, are shown in the figure and tabulated in Table~\ref{tab:trad}.
Because our fits are of two points to two parameters, there is no goodness-of-fit criterion to invoke:
The data cannot express a preference between the models.

The extrapolations to $a/L=0$ show a (mostly) downward trend compared to any individual $\sigma_P$.
The resulting error bars are of course larger but $\gamma_m$ never exceeds 0.5 for any of the couplings examined.
We defer further discussion of this sort until we reach the results of our all-volume analysis below.

\begin{figure*}
\begin{center}
\includegraphics[width=.95\columnwidth,clip]{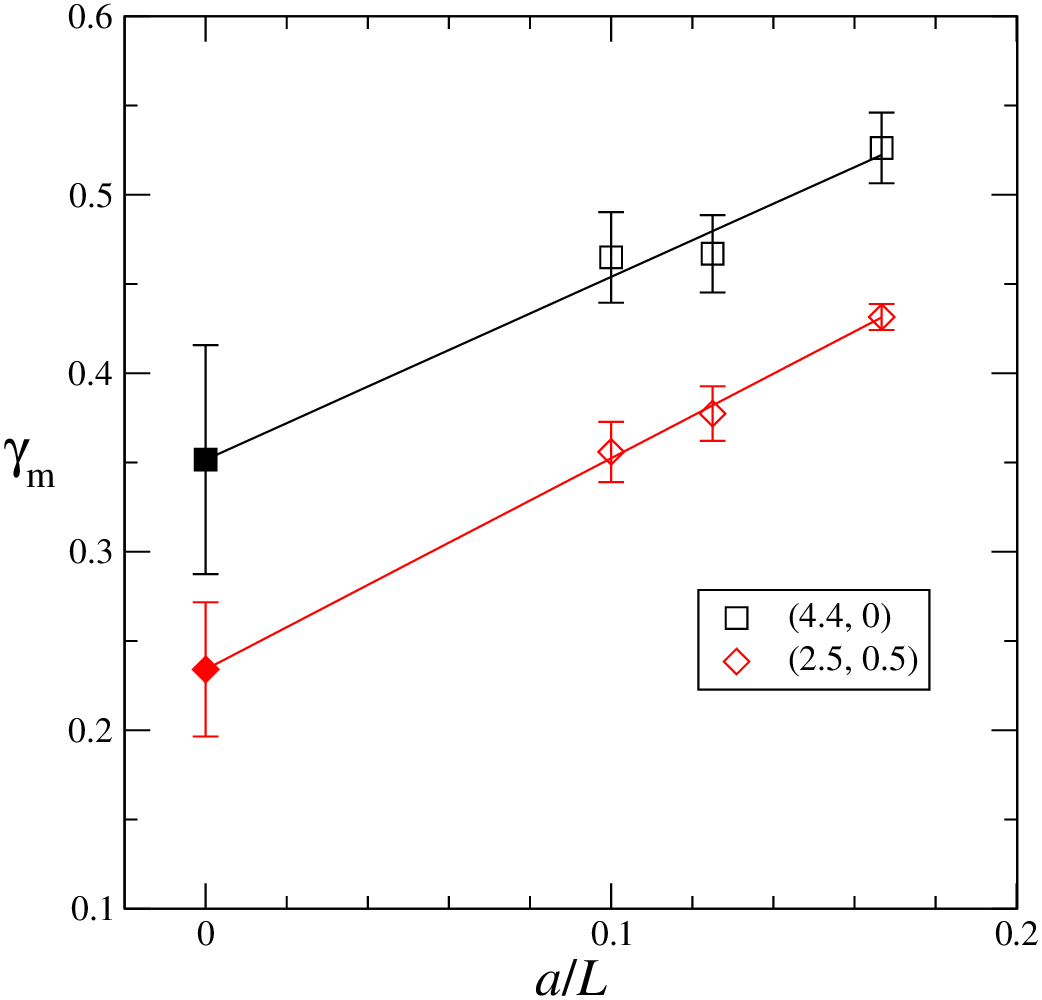}\hfill 
\includegraphics[width=.95\columnwidth,clip]{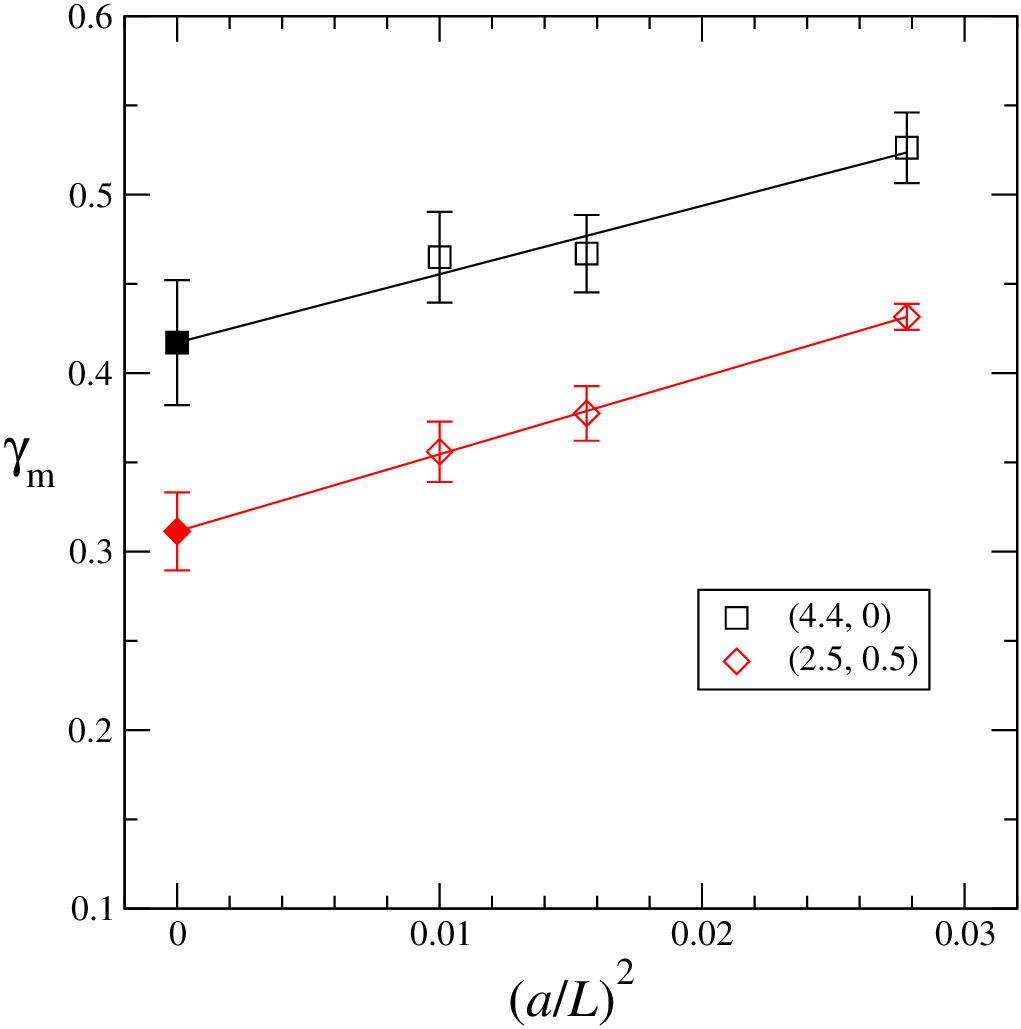}
\end{center}
\caption{$\gamma_m$ from fits to three pairs of lattices $(L,L')=(6a,12a)$, $(8a,14a)$, and $(10a,16a)$, plotted
versus $a/L$ (left) and versus $(a/L)^2$ (right). The extrapolated values are shown at the origin.
\label{fig:tradzp2}}
\end{figure*}

In  principle, we can judge the quality of the fits if we add data points, that is, more lattice volumes.
To this end we have run simulations for lattices of size $L=10a$ and~$14a$ at two of the couplings, $\beta=2.5$ and~4.4.
This gives us three independent pairs of volumes from which to calculate a step scaling ratio.
We use the pairs $L/a=(6,12)$, (8,14), and (10,16) to calculate $\sigma_P(g^2,s)$ for
$s=2$, 7/4, and~8/5, respectively, and estimate
\bee
\gamma_m(L)=\frac{\log\sigma_P(g^2,s)}{\log s}
\ee
as a function of the smaller $L$ of each pair.
Fig.~\ref{fig:tradzp2} shows the result, together with
the extrapolations to infinite $L$---now two-parameter fits to three points, which are also tabulated in Table~\ref{tab:trad}.
The quality of the fits is good in all cases, so (unfortunately) the data do not express a choice
between $1/L$ and $1/L^2$ corrections.

\begin{table}
\begin{center}
\begin{ruledtabular}
\begin{tabular}{cclc}
$\beta$ & $L$ pairs & \multicolumn2c{$\gamma_m$} \\
\cline{3-4}
&& \text{linear fit} & \text{quadratic fit} \\
\hline
4.4 & (6,12),(8,16)         & 0.37(10)&  0.44(5) \\
4.6 & (6,12),(8,16)         & 0.27(7) &  0.36(4) \\
2.0 & (6,12),(8,16)         & 0.43(7) &  0.43(4) \\
2.5 & (6,12),(8,16)         &  0.26(5)&  0.33(3) \\[2pt]
4.4 & (6,12),(8,14),(10,16) & 0.35(6) &  0.42(4) \\
2.5 & (6,12),(8,14),(10,16) & 0.23(4) &  0.31(3) \\
\end{tabular}
\end{ruledtabular}
\end{center}
\caption{``Traditional'' extrapolations of $\gamma_m$ from the pairwise step scaling function $\sigma_P$. 
\label{tab:trad}}
\end{table}

\subsubsection{All-volume extrapolations}

Now we present a procedure for using data from all volumes simultaneously to arrive at the extrapolated $\gamma_m$ at each value of the bare coupling.
The most straightforward approach would be to take the measurements of $Z_P$ and fit them to 
\bee
\log Z_P(L)= A + \gamma_m \log(a/L) + c\frac aL
\label{eq:plin}
\ee
or to
\bee
\log Z_P(L)= A + \gamma_m \log(a/L) +c\left(\frac aL\right)^2.
\label{eq:pquad}
\ee
We find, however, that our data do not distinguish well among the basis functions in these fits and hence the determinations of the parameters are unstable.
We therefore adopt an alternative approach to estimation of the discretization error.

For each bare coupling $\beta$ we have measured $Z_P$ on a set of volumes
$L_1<L_2<\ldots<L_N$.
The comparison in Fig.~\ref{fig:gamma} shows that
the running of the coupling is slow enough to be neglected
over the range $L_1 \le L \le L_N$.
The only obstruction to extracting $\gamma$
from a fit to \Eq{linfitZ} is then discretization error.
Since this error is most pronounced at the smallest $L$'s,
we can improve our estimate of the true $\gamma$ by successively
dropping the smaller volumes.  This observation is the basis for
the following continuum extrapolation.

Let $c_0^{(n)},c_1^{(n)}$ be the parameters of the linear fit (\ref{linfitZ}) where only the largest volumes $L_n,L_{n+1},\ldots,L_N$ are retained.
The slope $c_1^{(n)}$ is an estimate for $-\gamma$ which we take to be a function of
$a/L_n$.  For example, if $\lbrace L_1,\ldots,L_4\rbrace=\lbrace6a,8a,10a,12a\rbrace$,
a fit using all volumes gives an estimate $c_1^{(1)}$ of $-\gamma$ for
$a/L = 1/6$; dropping the smallest volume gives the next
estimate $c_1^{(2)}$ for $a/L = 1/8$; and dropping the two smallest volumes
gives the estimate $c_1^{(3)}$ for $a/L = 1/10$.
\begin{figure*}
\begin{center}
 \includegraphics[width=.95\columnwidth,clip]{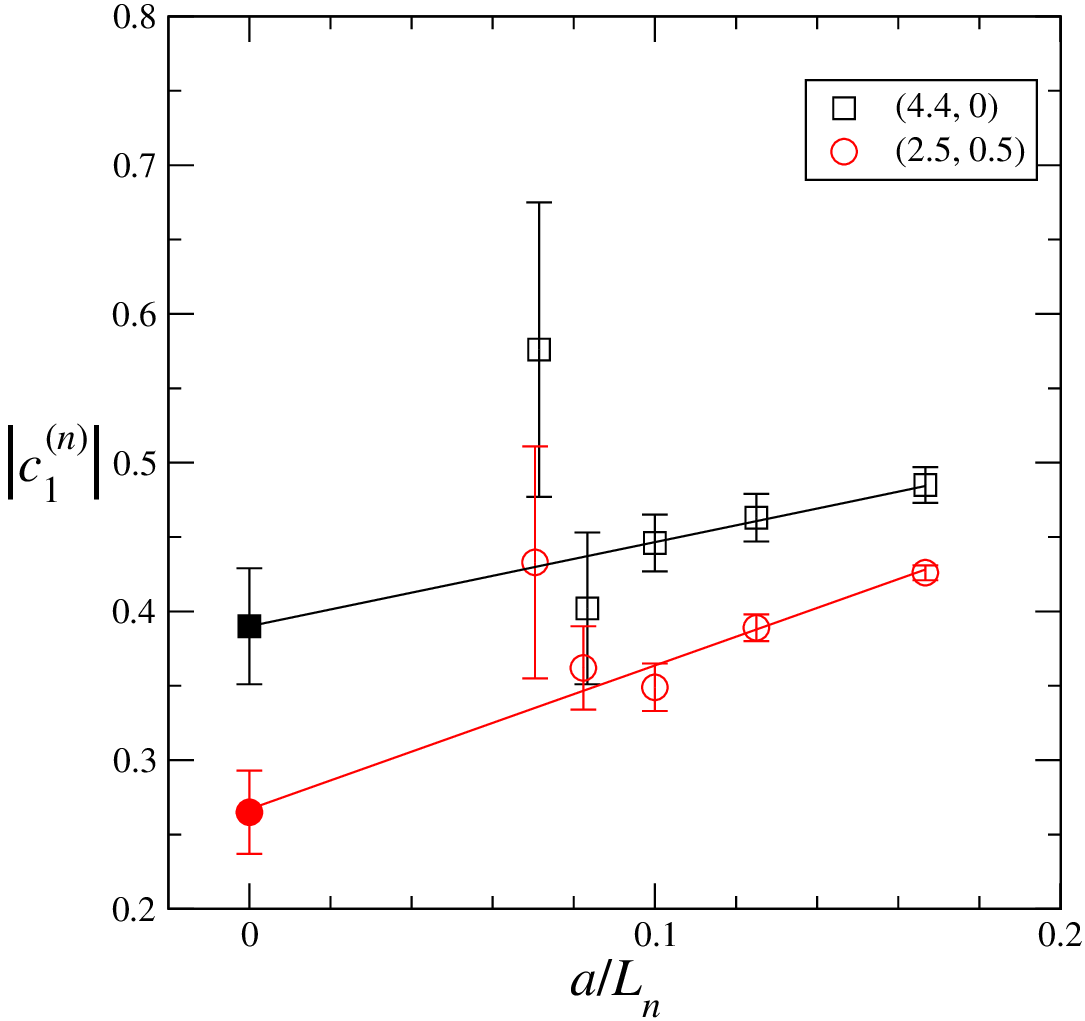}\hfill 
 \includegraphics[width=.95\columnwidth,clip]{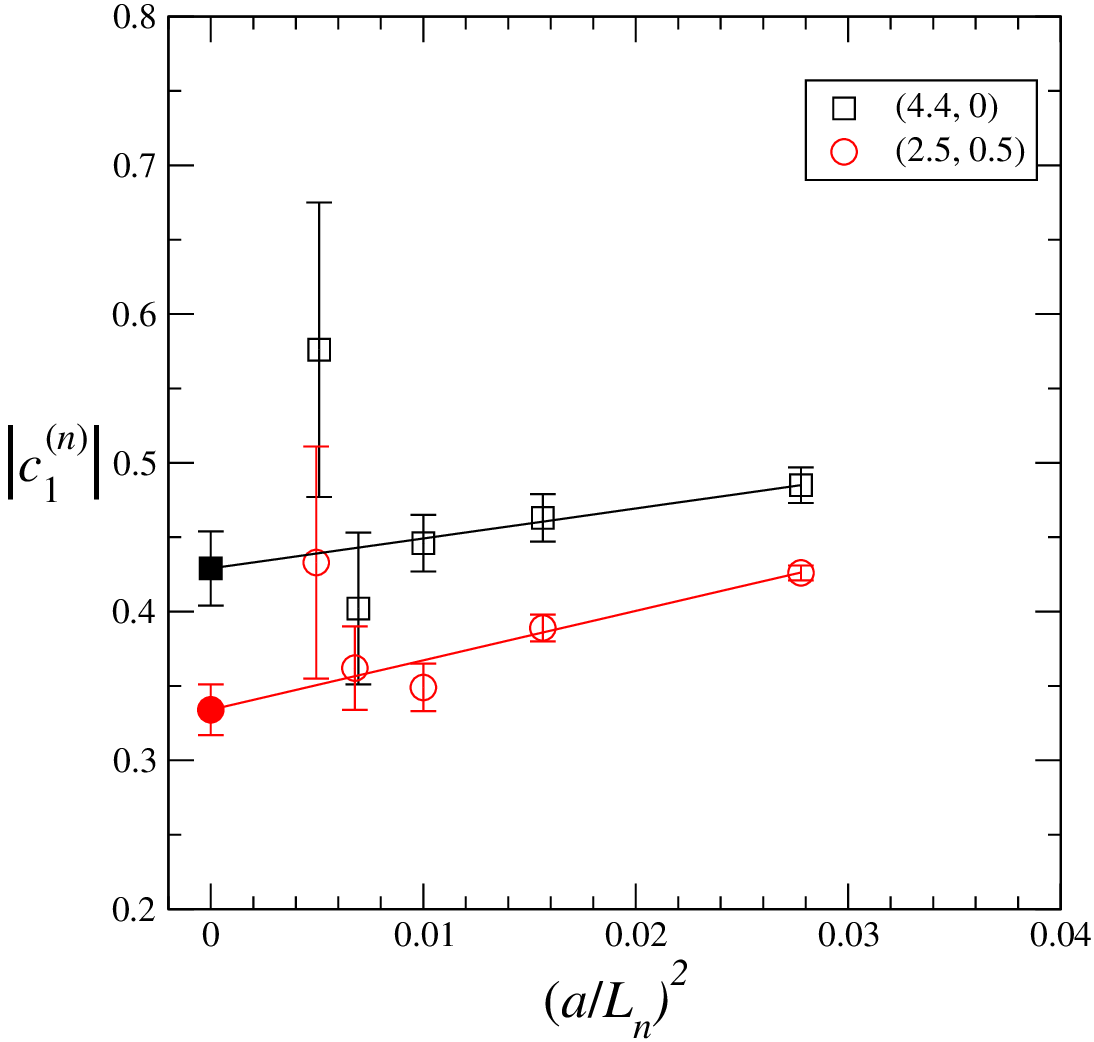}
\end{center}
\caption{Examples of linear and quadratic continuum extrapolations of the mass anomalous dimension $\gamma(g^2)$ derived from all-volume extrapolations.
Values of $(\beta,\beta_6)$ are as shown.
\label{fig:zp_dropmore}}
\end{figure*}
Finally we extrapolate these estimates to the continuum limit, $a/L\to 0$, by fitting to either \Eq{l1a1fit} or \Eq{l1a2fit}.

When performing the continuum extrapolation, we must take into
account that the results $c_1^{(n)}$ of the successive fits are correlated.
For the example above, with a maximum of two volumes dropped, it can be shown that the covariance
matrix ${\cal C}_{mn} = \text{cov}\left(c_1^{(n)},c_1^{(m)}\right)$ is
\bee
  {\cal C} = \left(\begin{array}{ccc}
               \D^{(1)} & \D^{(1)} & \D^{(1)} \\
               \D^{(1)} & \D^{(2)} & \D^{(2)} \\
               \D^{(1)} & \D^{(2)} & \D^{(3)}
               \end{array}\right) \ ,
\label{cov3x3}
\ee
where $\D^{(n)} = \sigma^2\left(c_1^{(n)}\right)$ is the variance of each $c_1^{(n)}$.

We show in Fig.~\ref{fig:zp_dropmore} the linear and quadratic continuum extrapolations for $\beta=2.5$ and~4.4, where we can work from six volumes.
In each case the fit has very low $\chi^2$ as long as we exclude the final (leftmost) data point, which is based on $L/a=14$ and~16 only.
Since its error bar is large, omitting this point from the fit leaves the extrapolation unchanged.
This is a feature of all the fits in the all-volume extrapolations.

\begin{figure}
\begin{center}
 \includegraphics[width=.8\columnwidth,clip]{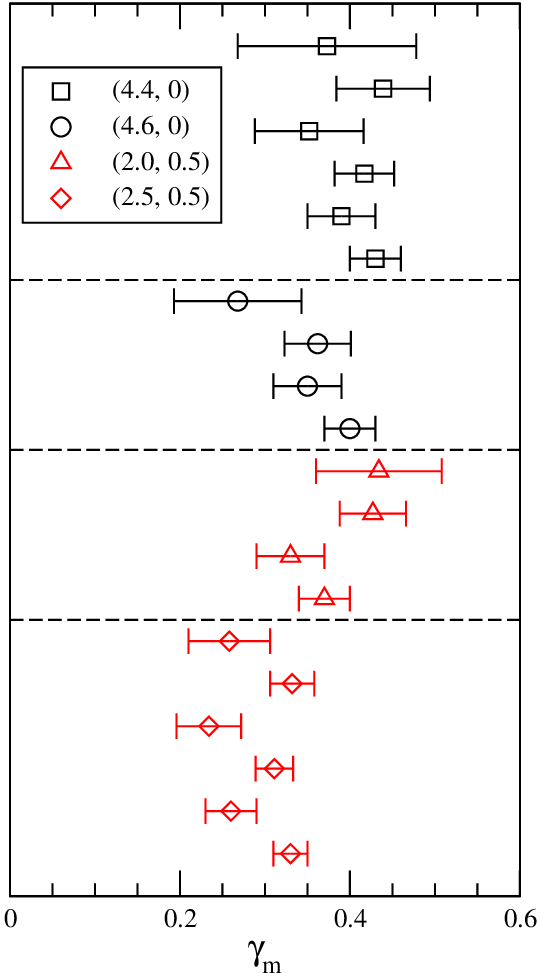}
\end{center}
\caption{Comparison of the results of all continuum extrapolations for the anomalous dimension at four bare couplings.
For each coupling, we plot in sequence, top to bottom: two-pair extrapolations---linear and quadratic (Fig.~\ref{fig:tradzp}); three-pair extrapolations (where available)---linear and quadratic (Fig.~\ref{fig:tradzp2}); all-volume extrapolations---linear and quadratic (Table~\ref{tab:scssv} and Fig.~\ref{fig:zp_dropmore}).
\label{fig:gamma_strip}}
\end{figure}

\begin{figure}
\begin{center}
 \includegraphics[width=.95\columnwidth,clip]{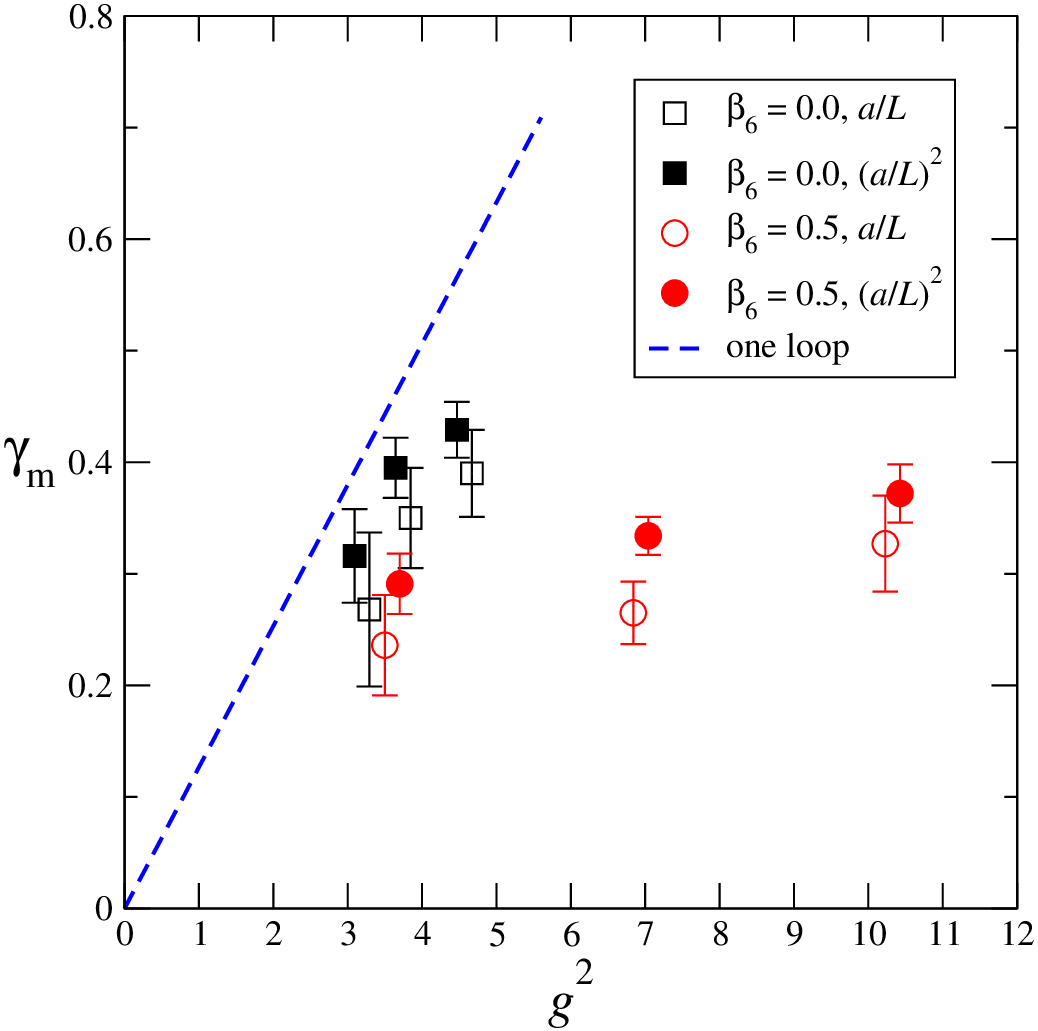}
\end{center}
\caption{Continuum limit of the mass anomalous dimension $\gamma(g^2)$ derived from all-volume extrapolations (see Table~\ref{tab:scssv}, and compare Fig.~\ref{fig:gamma}).
The squares are from the $\beta_6=0$ data while the circles are from $\beta_6=0.5$.
Empty symbols assume discretization errors proportional to $a/L$ while filled symbols assume $(a/L)^2$.
The filled symbols have been slightly displaced horizontally.
\label{fig:cont_gamma}}
\end{figure}

To give an idea of the stability of our fit procedures, we compare them all (as available) in Fig.~\ref{fig:gamma_strip}.
We display the results of the all-volume extrapolations for all bare couplings in Table~\ref{tab:scssv}, and we plot them in Fig.~\ref{fig:cont_gamma}.
Every point lies lower than the corresponding point in Fig.~\ref{fig:gamma}, which shows the result of the naive analysis.
Correction for discretization error has pushed our bound on $\gamma_m$ downward.
In addition, the gap between the two actions ($\beta_6=0$ vs.~0.5) has narrowed, for two reasons.
One is the larger error bars compared to Fig.~\ref{fig:gamma}; the other is the systematic uncertainty hovering over the choice between linear and quadratic continuum extrapolations.

\begin{table}
\begin{center}
\begin{ruledtabular}
\begin{tabular}{clcc}
$\beta$ & $L$ values & \multicolumn2c{$\gamma_m$} \\
\cline{3-4}
        &            & linear fit & quadratic fit \\
\hline
4.8 & 6,8,10,12       & 0.27(7) & 0.32(4) \\
4.6 & 6,8,12,16       & 0.35(4) & 0.40(3) \\
4.4 & 6,8,10,12,14,16 & 0.39(4) & 0.43(3) \\
3.5 & 6,8,10,12       & 0.24(5) & 0.29(3) \\
2.5 & 6,8,10,12,14,16 & 0.26(3) & 0.33(2) \\
2.0 & 6,8,10,12,16    & 0.33(4) & 0.37(3) \\
\end{tabular}
\end{ruledtabular}
\end{center}
\caption{Continuum limit of $\gamma_m$ from all-volume extrapolations. 
\label{tab:scssv}}
\end{table}

\subsection{Beta function}

\begin{figure}
\begin{center}
\includegraphics[width=.95\columnwidth,clip]{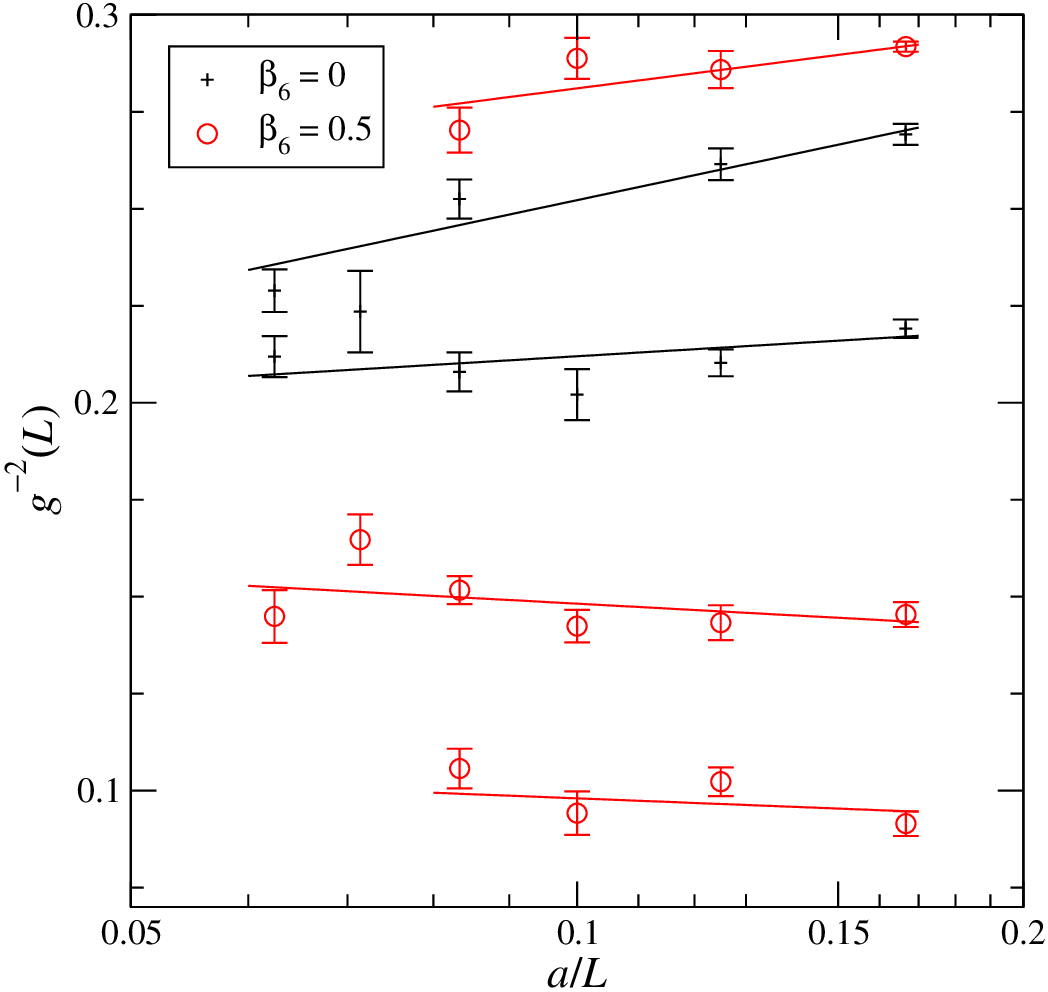}
\end{center}
\caption{
A blown-up view of Fig.~\ref{fig:couplings}, showing the strong-coupling portion of our data sets for the running coupling
  $1/g^2$.
\label{fig:coupl}}
\end{figure}

We would like to perform a similar analysis for the running coupling.
Unfortunately, the data do not
permit the extraction of extrapolated results carrying useful uncertainties. 
The reason why can quickly be 
seen from Fig.~\ref{fig:coupl}:
The data at each individual bare coupling are irregular functions of $L$. This may mean that the error bars for the individual data points are underestimated.
The SF observable (\ref{deta}) generally has a very long autocorrelation time. For lattice sizes $L\ge10a$, we determined an error bar by fitting 
four runs to a common constant, and collecting data until the 
$\chi^2$ per degree of freedom from this fit fell below about two.
This may not have been good enough. From a practical point of view, the pattern of our data makes analysis of the discretization errors quite difficult.

We shall treat only two data points: $\beta=2.0$ and~2.5, with $\beta_6=0.5$.
These are our strongest couplings, and as seen in Fig.~\ref{fig:beta} they hint at a positive beta function and hence an infrared fixed point.

At $\beta=2.0$ we have four volumes: $L=6a,8a,10a,12a$.
(We have discarded $L=16a$ as discussed above.)
For an almost traditional analysis we take them as two pairs, $(6a,10a)$ and $(8a,12a)$, and for each pair we calculate the discrete beta function (DBF),
\bee
B(u,s) = \frac1{g^2(sL)}-\frac1{g^2(L)}, \qquad u \equiv \frac1{g^2(L)}\ .
\label{DBF}
\ee
The scale factor $s$ is $5/3$ for the first pair and $3/2$ for the second pair.
The two DBFs can be combined if we rescale them according to
\bee
R(u,s)=\frac{B(u,s)}{\log s}\ .
\label{RDBF}
\ee
This rescaled DBF approximates the usual beta function $\tbeta(1/g^2)$ when the running is slow~\cite{DeGrand:2011qd}.
We can plot it against the smaller $L$ in each pair and extrapolate to $a/L=0$ either linearly or quadratically, much as we did for $\gamma_m$ above.

At $\beta=2.5$ we have six volumes.
Following our procedure for $\gamma_m$, we can first calculate the DBF for $s=2$, using the pairs $(L,2L)=(6a,12a)$ and $(8a,16a)$.  
Extrapolations to $a/L=0$ then follow.
Additionally, we calculate the rescaled DBF $R(u,s)$ for the three pairs
$(L,L')=(6a,12a)$, $(8a,14a)$, and~$(10a,16a)$, and extrapolate it to $a/L$=0.
Unfortunately, the irregularity of the data seen in Fig.~\ref{fig:coupl} renders the three-pair fit useless: Trying to fit a straight line through the three points gives an enormous $\chi^2$.
The two-pair fits, of course, have no degrees of freedom and hence no $\chi^2$.

At both couplings we can do all-volume extrapolations as for $Z_P$, {\em mutatis mutandis\/}.
We fit to \Eq{linfit} to obtain coefficients $c_0^{(n)},c_1^{(n)}$, where $n$ indicates the smallest volume $L_n$ retained in the fit.
Finally we extrapolate to $a/L=0$ either linearly or quadratically.

\begin{figure}
\begin{center}
 \includegraphics[width=.8\columnwidth,clip]{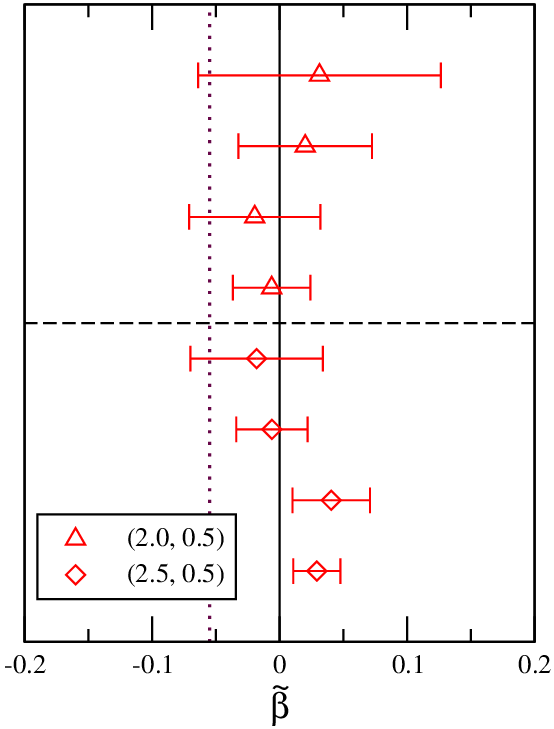}
\end{center}
\caption{Results of continuum extrapolations for the beta function at the two strongest couplings.
For each $(\beta,\beta_6)$ we plot in sequence, top to bottom, the two-pair extrapolations---linear and quadratic---and the all-volume extrapolations---linear and quadratic.
The vertical dotted line is the one-loop value.
\label{fig:beta_strip}}
\end{figure}

For $\beta=2.0$ the all-volume extrapolations give reasonable $\chi^2$ and improve on the pairwise extrapolations, as may be seen in Fig.~\ref{fig:beta_strip}.
Nonetheless, the error bar for each extrapolation has grown by a factor of 4--6 compared to the raw fit shown in Figs.~\ref{fig:couplings} and~\ref{fig:beta}, which wipes out any sign of a zero crossing.
For $\beta=2.5$ the all-volume extrapolations give positive results but, again, the error bar has grown by a factor of 4--6 and the central values lie barely $1\sigma$ above zero. 
We therefore cannot confirm the existence of an infrared fixed point.
It is quite possible that the beta function approaches zero as seen in Fig.~\ref{fig:beta} but then veers away, staying negative and leading to strong coupling at large distance.
This would be exactly the behavior conjectured for walking technicolor.

\section{Discussion \label{sec:summary}}

In this paper we have continued our study of the SU(3) gauge theory
with two flavors of sextet fermions.  A new term in the lattice action
allowed us to explore a much wider range of the renormalized coupling.
The old and new actions give rise to consistent results for the beta
function where they overlap.
For the mass anomalous dimension there are some disagreements;
we believe that the main source of the disagreement is the proximity
of the first-order phase transitions---a lattice artifact---in the old action.
The discrepancies are reduced when we extrapolate to the continuum limit.

We have developed a novel method to approach the continuum limit.
It begins with the observation that, in a truly conformal theory, there is only a single expansion
of lattice artifacts in powers of $a/L$.
This is in sharp contrast with QCD, which generates its own scale dynamically, and therefore requires
separate expansions for the discretization error and for finite-volume corrections.
In the slowly running theory, deviations from exact conformality can in principle be treated systematically. 
In order to overcome difficulties in resolving fit functions in the range of $a/L$ available to us,
we developed a method of extrapolation based on successively dropping the smallest lattices.
The differences between linear and quadratic extrapolations, as seen in Fig.~\ref{fig:gamma_strip}, are an indication of
our systematic error.

\subsection{An infrared fixed point?}

The question of the existence of an IRFP in this model has been addressed in the
literature in several ways.
A first hint comes from qualitative features of the phase diagram of the lattice theory~\cite{DeGrand:2010na,Svetitsky:2010zd}.
In QCD on a finite lattice, there is a confining phase for sufficiently strong coupling within which one may tune $\kappa\to\kappa_c$ in order to obtain a massless pion.
Then one can increase the size of the lattice while tuning towards weak coupling on the $\kappa_c(\beta)$ curve, thus approaching the continuum limit within the confining phase.
In the sextet theory, on the other hand, the confining phase is bounded by a first-order phase transition that does not permit definition of $\kappa_c$. 
It is not at all clear that a continuum limit in the confined phase would yield a theory with finite masses, let alone a chiral, massless limit.
The only alternative might be to begin in the weak-coupling, non-confining phase, where one could take a non-confining continuum limit.

This feature of the phase diagram was first seen in QCD with a large number of (color-triplet) flavors by Iwasaki et al.~\cite{Iwasaki:1991mr,Iwasaki:2003de}, and explored more recently in Ref.~\cite{Nagai:2009ip}.
This may be a sign of entry into the conformal window in these theories.
Turning to staggered fermions, the authors of Refs.~\cite{deForcrand:2012vh,deForcrand:2012sf} have recently noted the absence of spontaneous breaking of chiral symmetry in strong-coupling (triplet) QCD when the number of flavors is large. 

Even if one ignores the issue of a light pion,
one can ask whether any kind of continuum limit can be taken in the confining phase.
On a finite lattice, the first-order phase boundary at strong coupling connects to the boundary between strong-coupling and weak-coupling phases that may be interpreted as the finite-temperature phase transition.
In a confining theory, this transition must move towards weak coupling as the lattice is enlarged, in such a way as to give a finite transition temperature in the continuum limit.
In the theory with thin-link Wilson fermions, we found~\cite{DeGrand:2008kx} that this motion stalls near the weak-coupling $\kappa_c$ curve, as if the finite-temperature transition is avoiding the basin of attraction of an IRFP.
More detailed study of this question in the fat-link theory \cite{DeGrand:2010na} revealed a slow motion of the transition with lattice size; one would need to study larger lattices to see whether this motion eventually gives correct scaling in the continuum limit.
In fact, because the perturbative running of the coupling is so slow~\cite{DelDebbio:2008wb}, it will take enormous lattices to settle the matter.
Kogut and Sinclair \cite{Kogut:2010cz,Kogut:2011ty,Sinclair:2011ie,Sinclair:2012fa} have been studying the same question with staggered fermions; the results are, so far, similarly inconclusive.

Fodor {\em et al.}~\cite{Fodor:2011tw,Fodor:2012uu,Fodor:2012ty,Fodor:2012uw,Fodor:2012ni} 
have studied the staggered-fermion theory with extensive simulations on large volumes.
On each lattice, they use quark masses that are large enough that the volumes are effectively infinite.
They then test alternative scaling hypotheses for the mass spectrum, the chiral condensate, $f_\pi$, and the string tension as a function
of the quark mass.  Their analysis favors confinement over conformal physics,
meaning a beta function without an IRFP, which might behave as a walking theory.

As we have seen, our Schr\"odinger-functional results are not precise enough to nail down a fixed point.
Statistical fluctuations preclude unambiguous extrapolation to the continuum limit.
All we can say is that the beta function is smaller than the one-loop value, so that the theory runs slowly.

\subsection{The mass anomalous dimension}

As we have seen, $\gamma_m$ first follows
the one-loop curve and then at stronger couplings it levels off.
This is in line with what we found in the SU(2) and SU(4) theories~\cite{DeGrand:2011qd,DeGrand:2012qa}.
The fits of Sec.~\ref{sec:gamma}, which do not take lattice error into account, give a bound
\begin{equation}
  \gamma_m \alt 0.45 \,.
\label{boundgm}
\end{equation}
At weak couplings we did not carry out a continuum extrapolation.
At our strongest couplings, our analysis of discretization error (Sec.~\ref{sec:syst}) leads to larger error bars than our first analysis, but also to a general movement downwards so that our final result is the same bound (\ref{boundgm}).

The issue of scheme-dependence always arises at this point.
The value of $\gamma_m$ at a fixed point,
if there is one, is scheme-independent.
We can say more than this, however.
A bound like \Eq{boundgm} is evidently invariant under any redefinition $g\to g'(g)$, that is, it is entirely scheme-independent.
A change of scheme will change the functional dependence  of $\gamma$ on the  renormalized coupling, but if $\gamma$ is bounded (and even flat) over a wide range of bare couplings then it will stay so.

A successful model of walking (extended) technicolor must have
a slowly-varying coupling, but no infrared fixed point; it must also
have a large mass anomalous dimension, $\gamma_m \simeq 1$.
While our results do not rule out walking in the sextet theory,
the smallness of $\gamma_m$ makes this theory unsuitable for
walking technicolor.

\begin{acknowledgments}
We thank Ohad Raviv for assistance in the preparation of the appendix, and Sara Beck for a valuable discussion.
B.~S. and Y.~S. thank the University of Colorado for hospitality.
This work was supported in part by the Israel Science Foundation
under grant no.~423/09 and by the U.~S. Department of Energy.

The computations for this research were made possible in part by the U.~S. National Science Foundation through TeraGrid resources provided by (1) the University of Texas and (2) the National Institute for Computational Sciences (NICS) at the University of Tennessee, under grant number TG-PHY090023.
Additional computations were done on facilities of the USQCD Collaboration at Fermilab,
which are funded by the Office of Science of the U.~S. Department of Energy.
Our computer code is based on the publicly available package of the
 MILC collaboration~\cite{MILC}.
The code for hypercubic smearing was adapted from a program written by A.~Hasenfratz,
R.~Hoffmann and S.~Schaefer~\cite{Hasenfratz:2008ce}.
\end{acknowledgments}

\appendix*
\section{\label{app:oneloop} One loop analysis}

In the one-loop approximation the SF coupling,
defined in \Eq{deta}, takes the form
\bee
  \frac{1}{g^2(L)} = \frac{1}{g_0^2} + \Sigma_G(L) + 2 \Sigma_F(L) \ .
\label{Ksf1}
\ee
Here $g_0$ is the bare coupling from \Eq{g0eff},
and the one-loop contributions are $\Sigma_G$ from the gauge
and ghost fields%
\footnote{$\Sigma_G(L)$ has not been calculated for nHYP links.}
 and $\Sigma_F$ from each flavor of the sextet fermions.
Following the definitions%
\footnote{Our $\Sigma_F(L)$ is called $p_{1,1}$ in Ref.~\cite{Sint:1995ch}.}
 and methodology of Ref.~\cite{Sint:1995ch}, two groups \cite{Sint:2011gv,Sint:2012ae,Karavirta:2012qd,Karavirta:2012jx} have recently calculated the fermion piece $\Sigma_F$ for a number of gauge groups and representations
and for a variety of boundary conditions and background fields (as imposed in the SF method).
On general grounds, one can expand
\beea
  \Sigma_F(L) &=& r_0 + s_0\log(L/a) + r_1 \frac{a}{L}
  + s_1\frac{a}{L}\log(L/a)  \nonumber\\
  && + r_2 \frac{a^2}{L^2}
  + s_2\log(L/a) \frac{a^2}{L^2} + \cdots \ .
\label{expanddet}
\eea
The continuum limit consists of the first two terms on the right-hand side.
The coefficient $s_0$ of the logarithmic term is the universal
contribution of a fermion field to the one-loop beta function.
The constant $r_0$ depends on the choice of the background field, {\it i.e.},
it introduces scheme dependence.

All other terms on the right-hand side of \Eq{expanddet} constitute the
discretization error.  In the presence of a clover term with $c_{SW}=1$
we have $s_1=0$.  The remaining terms do not vanish in general.
The linear discretization error ($r_1$ term) can be cancelled
by adding a boundary counterterm for the gauge field.
As for the quadratic discretization error (the $r_2$ term), one can suppress
it by a judicious choice of the background field
\cite{Sint:2011gv,Sint:2012ae,Karavirta:2012qd}.
The outcome of adopting both measures is an improved one-loop behavior,
where $s_1=r_1=r_2=0$.

The above strategy is natural for QCD, where the bare coupling
is such that physics at the cutoff scale
is typically well inside the perturbative regime.
Here we are faced with a drastically
different situation.  Because of the slow running, in order to reach
the interesting strong values of the renormalized coupling, the bare coupling
must already be strong.  As a result, we can no longer rely
on perturbation theory at the quantitative level.  Suffice it to mention
that, in the vicinity of a two-loop fixed point, the two-loop
contributions must be as important as the one-loop!
Besides, the main thrust of the present paper is the calculation of the mass
anomalous dimension for sextet fermions,
for which no perturbative study of discretization
errors has been carried out.

The conclusion, obviously, is not that discretization errors are to be ignored;
as in any lattice simulation, a continuum extrapolation is mandatory.
The point is that the continuum extrapolation must be done
non-perturbatively.
This is what we have attempted, with mixed success, in Sec.~\ref{sec:syst}.

We now proceed to a discussion of $\Sigma_F$ for sextet fermions,
in order to get some idea of the magnitude of this piece of the discretization
error in the weak-coupling regime.
We have computed  $\Sigma_F$ by
following closely Appendix~A of Ref.~\cite{Sint:1995ch}.
Inputs to the calculation are the fermion twist phase, set to $\theta=\pi/5$,
and the classical background field,
which, in turn, is determined by the SF boundary data.
(nHYP smearing leaves the classical background field unchanged.)
The spatial links on each time boundary
are specified by three abelian phases $\phi_i=\phi_i(\eta)$, $i=1,2,3$,
which satisfy the constraint $\phi_1+\phi_2+\phi_3=0$.
The corresponding six phases of the sextet representation
are given by $\phi_{ij}=\phi_i+\phi_j$, with $1\le i\le j\le 3$.
$\Sigma_F$ is written as a sum over the lattice 3-momentum $\bf{p}$.
For each $\bf{p}$, the fermions' contribution can be represented as
$\partial/\partial\eta\,\tr\log M\Big|_{\eta=0}$,
where $M$ is a two-by-two matrix
that depends on $\bf{p}$ and on the boundary conditions \cite{Sint:1995ch}.

\begin{figure*}
\begin{center}
\includegraphics[width=.95\columnwidth,clip]{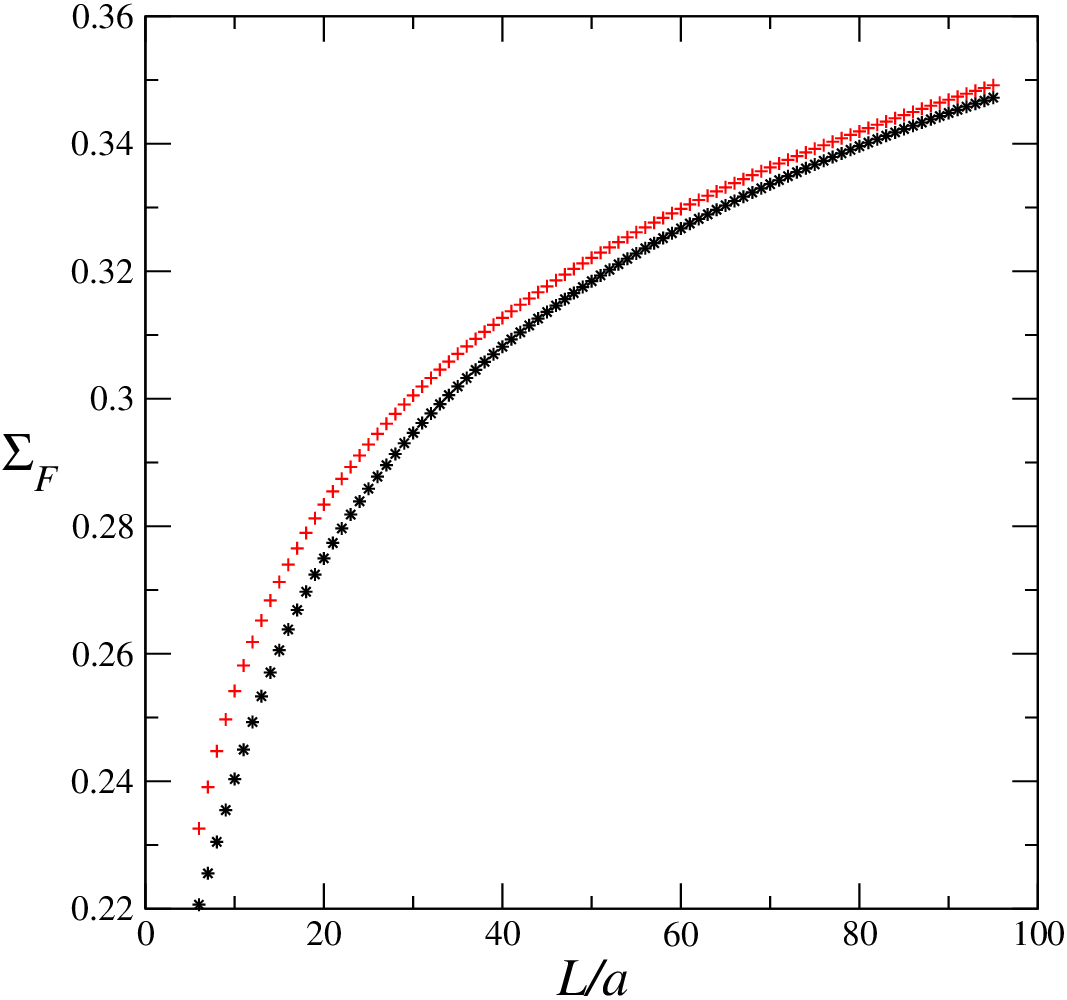}\hfill 
\includegraphics[width=.95\columnwidth,clip]{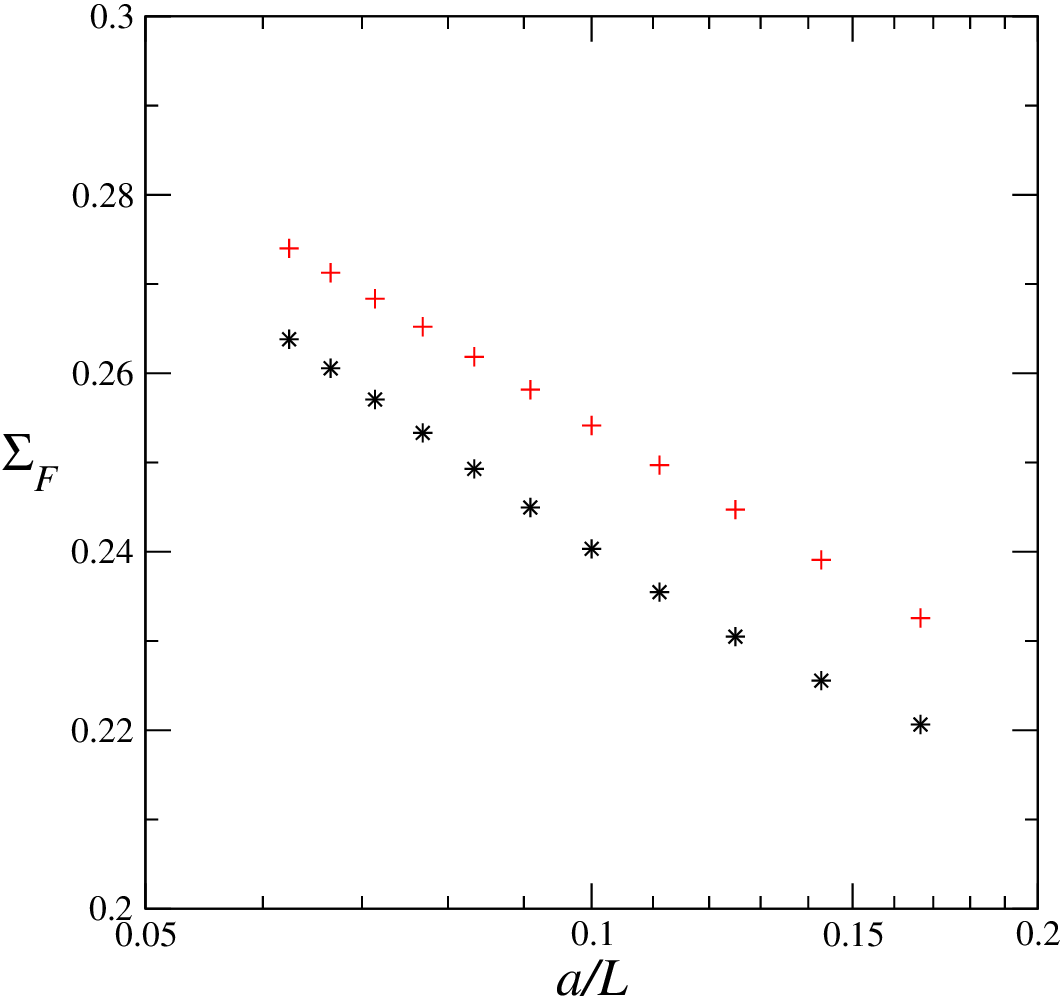}
\end{center}
\caption{The sextet fermions' contribution to the one-loop SF coupling as a function of lattice size.
The exact result (stars) is compared to the continuum formula (crosses) wherein the arbitrary constant $r_0$ comes from a fit as discussed in the text.
The second plot shows only $6a\le L\le16a$, for comparison with Fig.~\ref{fig:couplings}.
\label{fig:oneloop}}
\end{figure*}
We plot $\Sigma_F$ as a function
of $L$ in Fig.~\ref{fig:oneloop}.  For comparison, we also show the continuum result
$\widetilde\Sigma_F = r_0 + 5/(12\pi^2)\log L$,
where the scheme-dependent
constant $r_0$ has been extracted from a fit of $\Sigma_F$ to the six terms shown in \Eq{expanddet}.
In this fit, the coefficient $s_0$ is fixed at the continuum value and $s_1$ is fixed to zero (see above).
The other coefficients are listed in Table~\ref{table:sigfit}.

\begin{table}[b]
\caption{Free coefficients in the fit (\ref{expanddet}).}
\begin{center}
\begin{tabular}{dd}
\hline\hline
r_0 & 0.1569\\
r_1 & -0.1914\\
r_2 & 0.4169\\
s_2 & 0.0086 \\
\hline\hline
\end{tabular}
\end{center}
\label{table:sigfit}
\end{table}

In order to get an idea of the magnitude of discretization errors
in the beta function, we plot $\Sigma_F$ against $\log (a/L)$ in the second panel of Fig.~\ref{fig:oneloop};  here we limit the range of $L$
to $6a\le L \le 16a$ to match our numerical simulations.
This plot is analogous to Fig.~\ref{fig:couplings} (but note that here
we display only the contribution of a single sextet fermion).
The continuum result now shows up as a straight line,
whereas the lattice result exhibits deviations from linearity.
It can be seen that much of the discrepancy between the continuum and lattice
results can be accounted for by an additive constant.
As noted above, a constant shift $r_0$ does not represent
a discretization error, but, rather, scheme dependence.
The discretization error in the beta function will appear as a difference in the slopes of the two lines.
Using the method of least squares to estimate
the average slope of the lattice data we find the value 0.0446,
which amounts to a 5\% deviation from the correct continuum value
$5/(12\pi^2) \simeq 0.0422$.  Multiplying by two for the number of flavors,
we conclude that the fermions' contribution to the discretization error
in the beta function is in absolute terms roughly 0.005.
A glance at Fig.~\ref{fig:beta} shows that this is a less than half
the statistical error plotted for the weak-coupling data points.

The smallness of the deviation in the slope is connected to the coefficients
$r_1$ and $r_2$ of the linear and quadratic lattice corrections.
The two coefficients turn out to have opposite sign and magnitudes that cause their contributions to the slope to cancel.
Thus one cannot really justify an expansion in powers of $a/L$ in extrapolating to the continuum limit from our data.
As we have stressed above, any given extrapolation is a model.


\end{document}